\documentclass[swedish,english,5p]{elsarticle}
\usepackage[latin9]{inputenc}
\usepackage{babel}
\usepackage{amsmath}
\usepackage{graphicx}
\usepackage{color}
\usepackage[unicode=true,
 bookmarks=false,
 breaklinks=false,pdfborder={0 0 1},backref=section,colorlinks=false]
 {hyperref}

\makeatletter
\newdefinition{rmk}{Remark}
\newproof{pf}{Proof}

\makeatother

\begin{document}
\begin{frontmatter}

\title{A Bayesian Heteroscedastic GLM with Application to fMRI Data with
Motion Spikes}

\author[mymainaddress,mysecondaryaddress,mythirdaddress]{Anders Eklund\corref{mycorrespondingauthor}}

\author[myfourthaddress]{Martin A. Lindquist}

\author[mymainaddress]{Mattias Villani{*}}

\cortext[Mattias Villani]{Corresponding author}

\ead{mattias.villani@liu.se}

\address[mymainaddress]{Division of Statistics \& Machine Learning, Department of Computer
and Information Science, Link\"{o}ping University, Link\"{o}ping, Sweden}

\address[mysecondaryaddress]{Division of Medical Informatics, Department of Biomedical Engineering,
Link\"{o}ping University, Link\"{o}ping, Sweden}

\address[mythirdaddress]{Center for Medical Image Science and Visualization (CMIV), Link\"{o}ping University, Link\"{o}ping, Sweden}

\address[myfourthaddress]{Department of Biostatistics, Johns Hopkins University, Baltimore,
USA}
\begin{abstract}
We propose a voxel-wise general linear model with autoregressive noise
and heteroscedastic noise innovations (GLMH) for analyzing functional
magnetic resonance imaging (fMRI) data. The model is analyzed from
a Bayesian perspective and has the benefit of automatically down-weighting
time points close to motion spikes in a data-driven manner. We develop
a highly efficient Markov Chain Monte Carlo (MCMC) algorithm that
allows for Bayesian variable selection among the regressors to model
both the mean (i.e., the design matrix) and variance. This makes it
possible to include a broad range of explanatory variables in both
the mean and variance (e.g., time trends, activation stimuli, head
motion parameters and their temporal derivatives), and to compute
the posterior probability of inclusion from the MCMC output. Variable
selection is also applied to the lags in the autoregressive noise
process, making it possible to infer the lag order from the data simultaneously
with all other model parameters. We use both simulated data and real
fMRI data from OpenfMRI to illustrate the importance of proper modeling
of heteroscedasticity in fMRI data analysis. Our results show that
the GLMH tends to detect more brain activity, compared to its homoscedastic
counterpart, by allowing the variance to change over time depending
on the degree of head motion. 
\end{abstract}
\end{frontmatter}

\section{Introduction}

Functional magnetic resonance imaging (fMRI) is a non-invasive technique
that has become the de facto standard for imaging human brain function
in both healthy and diseased populations. The standard approach for
analyzing fMRI data is to use the general linear model (GLM), proposed
by Friston et al. \citep{friston}. The standard GLM has been extremely
successful in a large number of empirical studies, but relies on a
number of assumptions, including linearity, independency, Gaussianity
and homoscedasticity (constant variance). Much work has been done
to relax the assumption of independent errors, and several alternative
noise models have been proposed \citep{Friston1999,Woolrich2001,Lund2006,lenoski,eklund2012Neuroimage}.
In addition, it has also been investigated whether results are improved
by using a Rician noise model \citep{gudbjartsson1995rician,AdrianMaitraRowe2013Ricean,noh2011rician,solo2007EMRician},
instead of a Gaussian. While heteroscedastic models exist for group
analyses \citep{Beckmann,flame,chen}, the homoscedasticity assumption
for single subject analysis has received little attention. \citet{nichols2003}
used the Cook-Weisberg test for homoscedasticity to detect problematic
voxels, but did not propose a heteroscedastic model to handle these.
\citet{Diedrichsen2005} claim that the homoscedasticity assumption
is often violated in practice due to head motion, and propose an algorithm
that estimates the noise variance separately at each time point. The
estimated variances are then used to perform weighted least squares
regression. The aim of this study is to further explore the appropriateness
of the homoscedasticity assumption for single subject fMRI analysis,
and evalute the effects of deviations from it.

\subsection{Is fMRI noise heteroscedastic?}

\begin{figure}
\centering\includegraphics[scale=0.45]{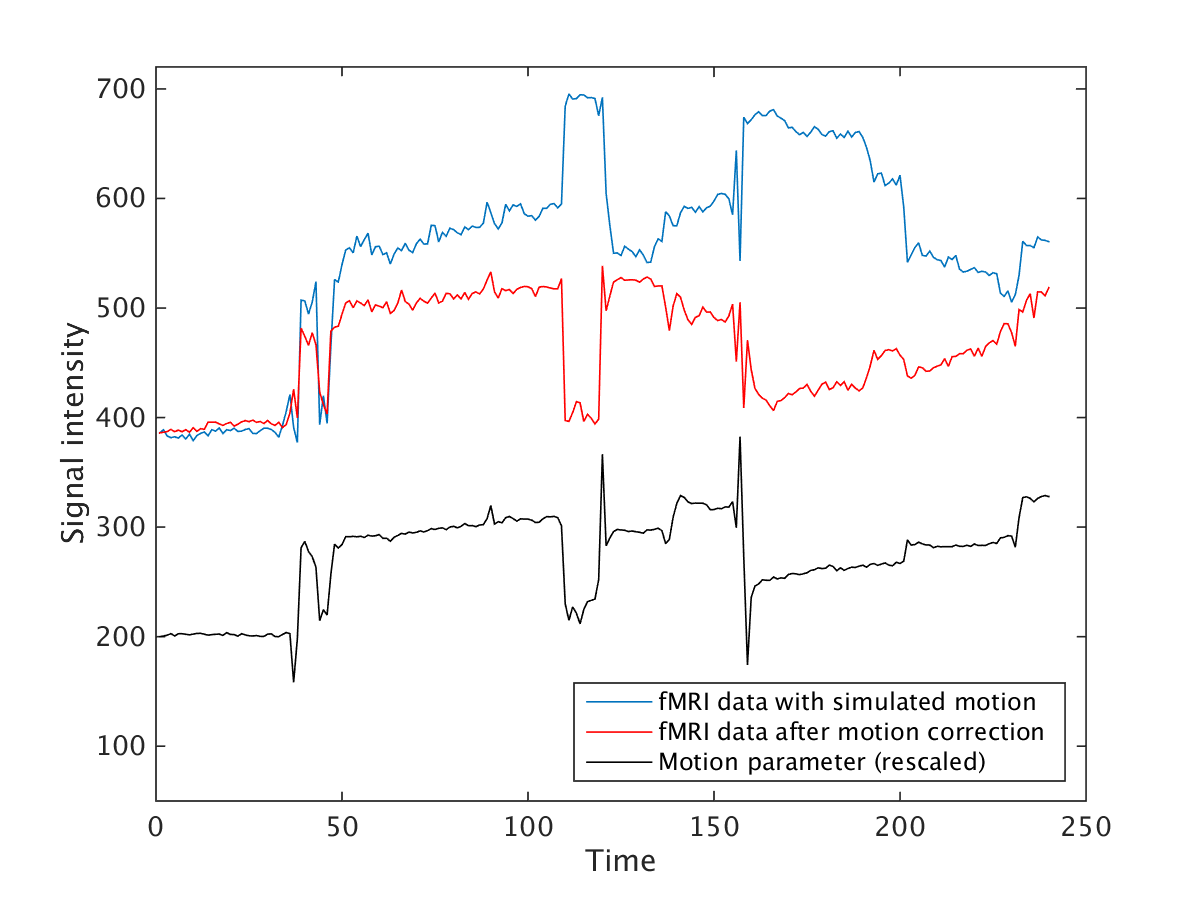}

\caption{ A single time series from a simulated fMRI dataset, before and after
motion correction. {\color{black} All the signal variation comes from simulated motion, and is due to interpolation artefacts.} Note that the motion corrected data still has a
high signal variance, which is correlated with the head motion.\label{fig:heterosim1}}
\end{figure}

\begin{figure}
\centering\includegraphics[scale=0.45]{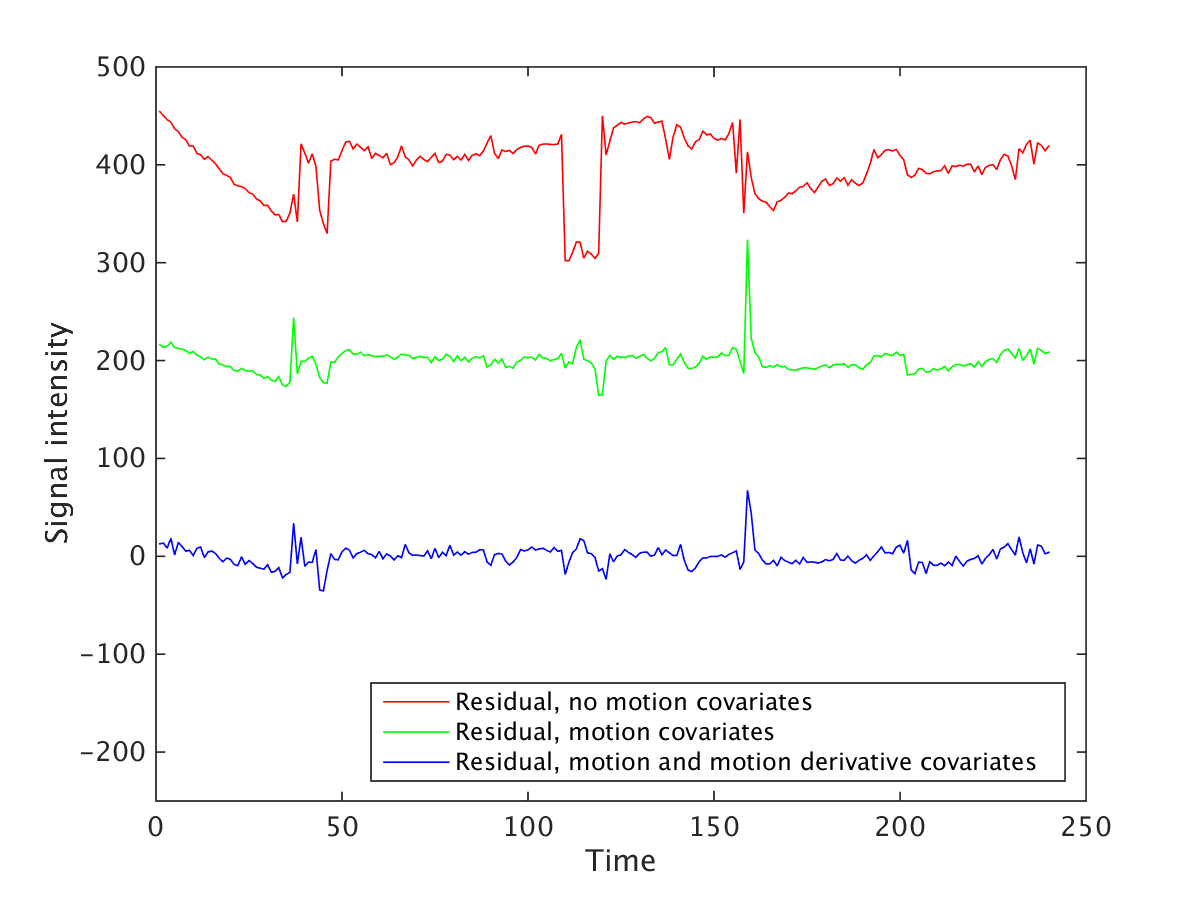}

\caption{Residual time series obtained after fitting models with three different
design matrices. The first design matrix only contains covariates
for the intercept and time trends (4 covariates in total). The second
design matrix also contains head motion covariates (10 covariates
in total), and the third design matrix also contains the temporal
derivative of the head motion (16 covariates in total). For visualization
purposes, a mean of 200 was added to the green residual, and a mean
of 400 was added to the red residual. Note that a motion spike is
still present in the green and the blue residual, making the noise
heteroscedastic. \label{fig:heterosim2}}
\end{figure}

Consider a simple simulation where actual head motion is applied to a single volume from {\color{black} a real} fMRI dataset, to generate a new 4D fMRI dataset where all the signal variation comes from simulated
motion. {\color{black} For each time point, the corresponding head motion parameters are used to translate and rotate the first volume in the dataset (using interpolation), and the transformed volume is saved as the volume for that specific time point.} Even if motion correction is applied to the {\color{black} simulated} dataset, the dataset will still contain motion related signal variation \citep{grootoonk}, due to the fact that the interpolation mixes voxels with low and high
signal intensity (especially at the edge of the brain, and at the
border between different tissue types). It is therefore common to
include the estimated head motion parameters in the design matrix,
to regress out any motion related variance that remains after the
motion correction, and to also account for spin-history artifacts.
It is also common to include the temporal derivative of the head motion
parameters, to better model motion spikes. Figure~\ref{fig:heterosim1}
shows a single time series from an fMRI dataset with simulated motion,
before and after motion correction, and one of the head motion parameters.
The selected voxel is at the border between white and gray matter.
Figure~\ref{fig:heterosim2} shows three residual time series calculated
using three different design matrices (and ordinary least squares
regression), the first containing only an intercept and time trends,
the second also containing motion covariates, and the third also containing
the temporal derivative of the head motion. It is clear that using
the estimated head motion as additional covariates removes most of
the motion related variance, but not all of it. The residual time
series still contain effects of a motion spike, which makes the noise
heteroscedastic. 

It should also be stressed that real fMRI data are far more complicated, for example due to the fact that each fMRI volume is sampled one slice at a time. {\color{black} Another problem is so-called 'spin-history' effects which alter the signal intensity of volumes following the motion spike, because the head motion changes the excitation state of the spins of the protons (thereby interrupting the steady state equilibrium). For this reason, a number of volumes after the motion spike should also be downweighted, and not only volumes during the motion spike.} 

\subsection{Modeling the heteroscedasticity}

We propose a Bayesian heteroscedastic extension of the GLM, which
uses covariates for both the mean and variance, and also incorporates
an autoregressive noise model. We develop highly efficient Markov
Chain Monte Carlo (MCMC) algorithms for simulating from the joint
posterior distribution of all model parameters. Allowing for heteroscedasticity,
where the noise variance is allowed to change over time, has the effect
of automatically discounting scans with large uncertainty when inferring
brain activity or connectivity. One way of thinking of this effect
is in terms of weighted least squares estimation, where the optimal
weights are learned from the data. 

\subsection{Is fMRI noise heteroscedastic in all voxels?}

\begin{figure}
\centering\includegraphics[scale=0.45]{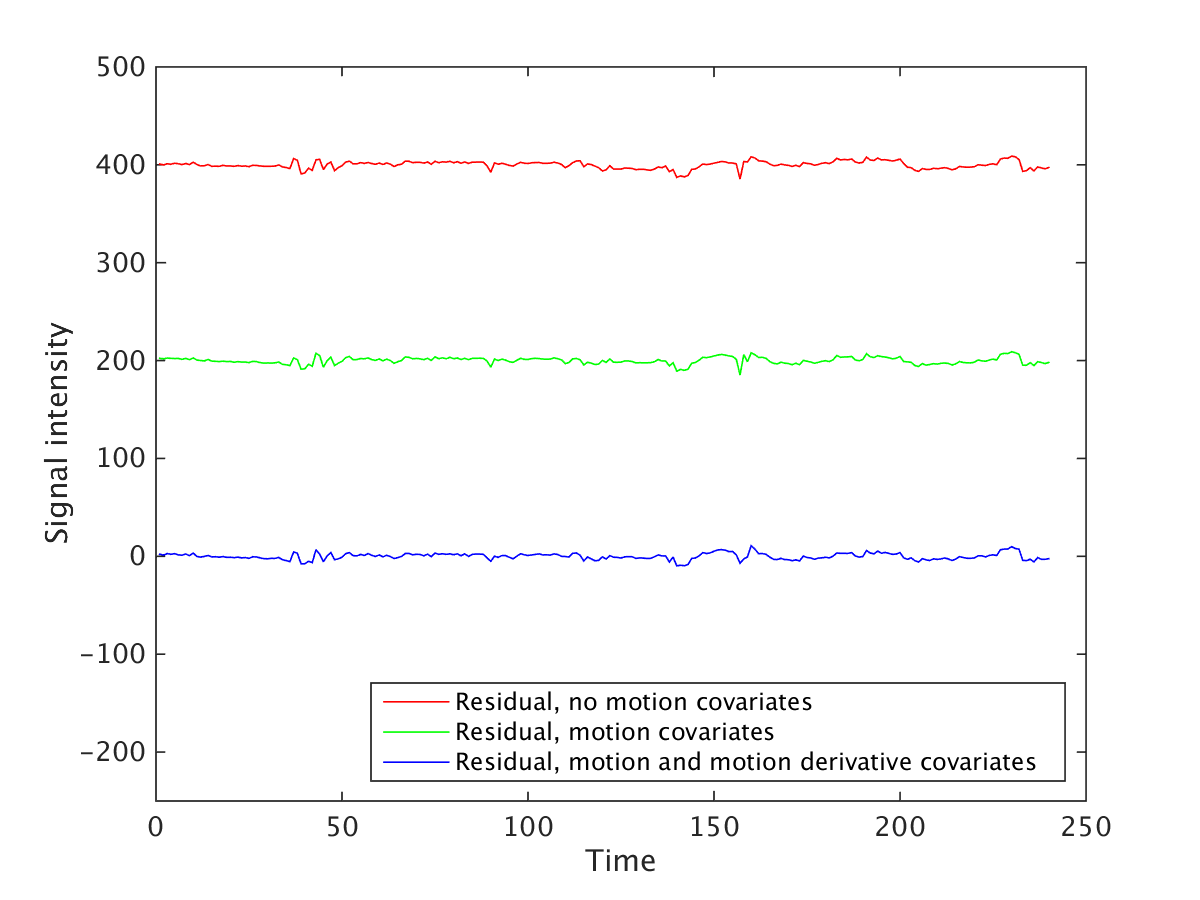}

\caption{Residual time series obtained after fitting models with three different
design matrices. The first design matrix only contains covariates
for intercept and time trends (4 covariates in total). The second
design matrix also contains head motion covariates (10 covariates
in total), and the third design matrix also contains the temporal
derivative of the head motion (16 covariates in total). For visualization
purposes, a mean of 200 was added to the green residual, and a mean
of 400 was added to the red residual. Note that the time series in
this voxel has a very low correlation with the simulated motion, and
the residuals are therefore homoscedastic.\label{fig:heterosim3}}
\end{figure}

Figure~\ref{fig:heterosim3} shows three residual time series for
a voxel in gray matter (close to the voxel shown in Figure~\ref{fig:heterosim2}).
Clearly, this voxel has a very low correlation with the simulated
motion, and the residuals are not heteroscedastic. It is therefore
not optimal to use the same weights in all voxels. Compared to the
work by \citet{Diedrichsen2005}, our Bayesian approach independently
estimates a heteroscedastic model for each voxel, instead of using
variance scaling parameters that are the same for all voxels. Furthermore,
\citet{Diedrichsen2005} used a fix autoregressive (AR) model for
the noise (AR(1) + white noise with the AR parameter fixed to 0.2,
as in the SPM software package), while we estimate an AR($k$) model
in each voxel. The fixed AR(1) model used by SPM has been shown to
perform poorly \citep{eklund2012Neuroimage}, especially for short
repetition times made possible with recently developed MR scanner
sequences.

Our Bayesian approach also differs from recently developed methods
used in the field, where scrubbing or censoring is used to remove
volumes with excessive head motion \citep{Satterthwaite,Power2014,siegel}.
Such approaches are ad hoc in the sense that an arbitrary motion threshold
first needs to be applied, to determine which volumes to remove or
censor. Another problem with these approaches is that they can significantly
alter the temporal structure of the data. 

\subsection{Variable selection}

It can be difficult to determine which variables to include in the
design matrix (i.e., the mean function) of the GLM, including those
that capture scanner drift, or residual head movement effects after
motion correction. It can be even more difficult to choose the appropriate
explanatory variables to use in the variance function. For this reason
we introduce variable selection priors in both the mean and variance
function, which has the effect of automatically determining the set
of explanatory variables; more precisely, we obtain the posterior
inclusion probability for each of the candidate variables and the
posterior distribution of their effect sizes from a single MCMC run.
In addition, we have a third variable selection prior acting on the
lags of the AR noise process which allows us to estimate the model
order of the AR process directly from the data. This aspect is particularly
important for high (sub-second) temporal resolution data. Our analysis
here is massively univariate without modeling spatial dependencies,
however we discuss possible extensions to spatial models in the Discussion.

\section{GLM with heteroscedastic autoregressive noise}

We propose the following voxel-wise GLM with heteroscedastic noise
innovations (GLMH) for blood oxygenation level dependent (BOLD) time
series:

\begin{align}
y_{t} & =\mathbf{x}_{t}^{T} \boldsymbol{\beta}+u_{t}\nonumber \\
u_{t} & =\rho_{1}u_{t-1}+...+\rho_{k}u_{t-k}+\exp(\mathbf{z}_{t}^{T} \boldsymbol{\gamma} /2)\cdot\varepsilon_{t},\;t=1,...,T,
\end{align}
where $y_{t}$ is the observed fMRI signal at time $t$, $\mathbf{x}_{t}$
is a vector with $p$ covariates for modeling the mean, $\varepsilon_{t}$
is zero mean Gaussian white noise with unit variance, and $\mathbf{z}_{t}$
is a vector of $q$ covariates for modeling the variance of the heteroscedastic
noise innovations as $\ln\sigma_{t}^{2}=\mathbf{z}_{t}^{T} \boldsymbol{\gamma}$. {\color{black} The logarithm of the variance is modelled as a linear regression, to enable unrestricted estimation of $\boldsymbol{ \gamma}$ while still guaranteeing a positive variance.} Note that we are here using the logarithmic link function for the
variance, but our methodology is applicable to any invertible and
twice-differentiable link function. The GLMH model introduces heteroscedasticity
through noise innovations with the effect that a large variance at
time $t$ is likely to generate a large innovation in the $u_{t}$
equation, which is propagated through the autoregressive structure.
The effect is that the noise remains large in subsequent scans, which
is desireable as it has been shown that motion related signal changes
can persist more than 10 seconds after motion ceases \citep{Power2014} {\color{black} (for example due to spin-history effects, as mentioned in the Introduction)}.

\begin{figure}
\includegraphics[scale=0.45]{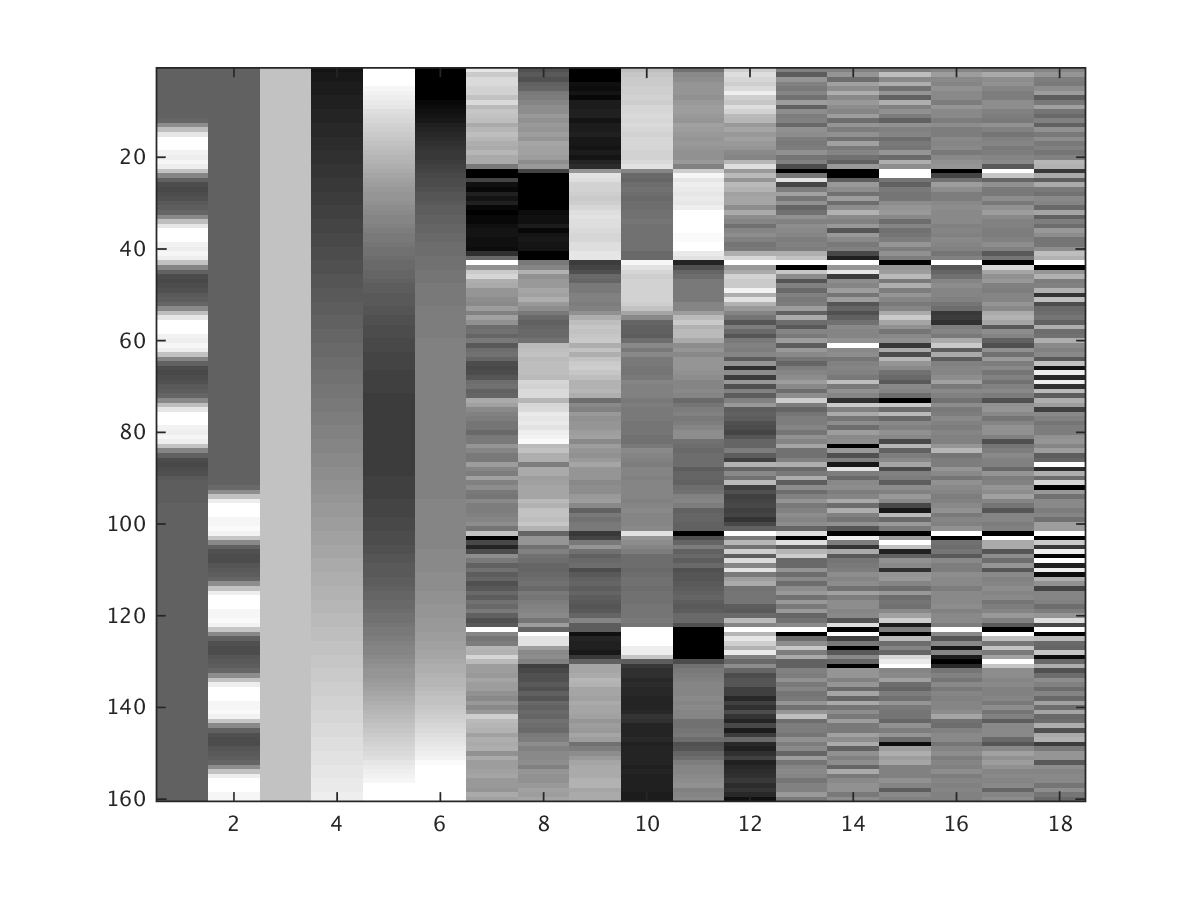}
\includegraphics[scale=0.45]{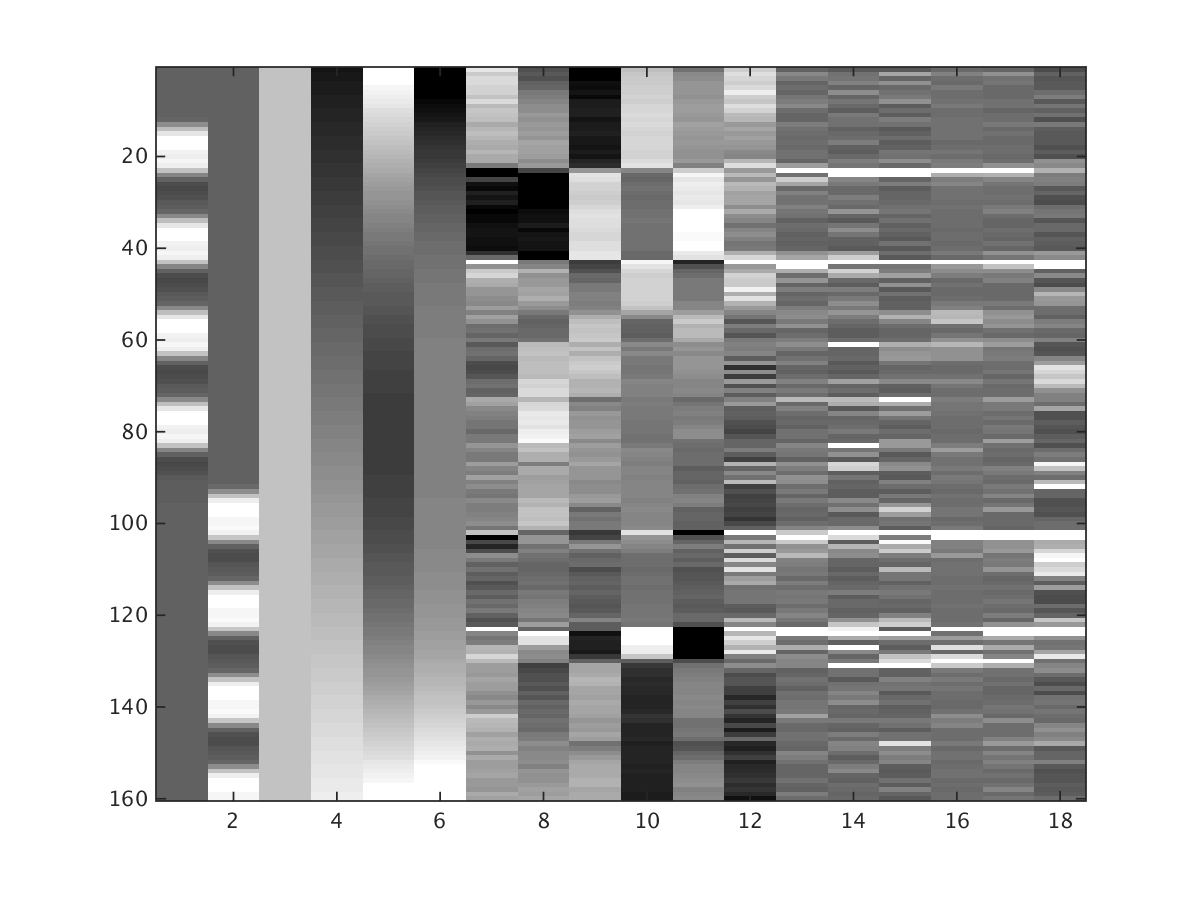}
\caption{{\color{black} An example of $\mathbf{X}$ (top) and $\mathbf{Z}$ (bottom) for a subject with several motion spikes. The data consists of 160 time points, and 18 covariates are here used to model both the mean ($\mathbf{X}$) and the variance ($\mathbf{Z}$). The first two covariates (from the left) represent two different tasks, the following four covariates represent intercept, linear trend, quadratic trend and cubic trend, the following six covariates represent the estimated head motion, and the last six covariates represent the temporal derivative of the head motion. The only difference between  $\mathbf{X}$ and $\mathbf{Z}$ is that the absolute value of the temporal derivative is used for $\mathbf{Z}$, as a motion spike should always lead to an increase of the variance. All covariates (except the intercept) are standardized to have zero mean and unit variance, which often leads to a better convergence of the MCMC chain.} \label{fig:XZexample}}
\end{figure}

Let $\mathbf{y}=(y_{1},...,y_{T})^{T}$ be a $T$-dimensional vector
consisting of observed fMRI signals at a specific voxel and define
$u$ and $\varepsilon$ analogously. Also, define $\mathbf{X}=(\mathbf{x}_{1},...,\mathbf{x}_{T})^{T}$
and $\mathbf{Z}=(\mathbf{z}_{1},...,\mathbf{z}_{T})^{T}$ to be $T\times p$
and $T\times q$ matrices consisting of covariates. Further, let $\boldsymbol{\rho}=(\rho_{1},...,\rho_{k})^{T}$.
The GLMH model can then be written as follows: 
\begin{align}
\mathbf{y} & =\mathbf{X} \boldsymbol{\beta}+\mathbf{u}\nonumber \\
\mathbf{u} & =\mathbf{U} \boldsymbol{\rho}+\mathrm{Diag}\left(\exp\left(\mathbf{Z} \boldsymbol{\gamma}/2\right)\right) \boldsymbol{\varepsilon},\label{eq:HIGLMfullSample}
\end{align}
 where $\mathbf{U}$ is a $T\times k$ matrix consisting of lagged
values of $\mathbf{u}$, assuming that $k$ pre-sample observations
are available. {\color{black} Figure~\ref{fig:XZexample} shows an example of $\mathbf{X}$ and $\mathbf{Z}$ for a subject with several motion spikes.  } 

\section{Bayesian inference\label{sec:Bayesian-Inference}}

We begin by defining the binary indicators $\mathcal{I}_{\beta}$,
$\mathcal{I}_{\gamma}$, and $\mathcal{I}_{\rho}$, which are used
for variable selection purposes. Here $\mathcal{I}_{\beta}$ is a
$p\times1$ vector whose jth element takes the value 1 if $_{j}$
is non-zero and 0 otherwise. The indicators $\mathcal{I}_{\gamma}$
and $\mathcal{I}_{\rho}$ are defined analogously. We take a Bayesian
approach with the aim of computing the joint posterior distribution
$p(\boldsymbol{\beta},\boldsymbol{\gamma},\boldsymbol{\rho},\mathcal{I}_{\beta},\mathcal{I}_{\gamma},\mathcal{I}_{\rho}|\mathbf{y},\mathbf{X},\mathbf{Z})$.
This distribution is intractable and we use Metropolis-within-Gibbs
sampling \citep{mcmchandbook} to generate draws from the joint posterior.
The algorithm iterates between the following three full conditional
posteriors:
\begin{enumerate}
\item $(\boldsymbol{\beta},\mathcal{I}_{\beta})\vert\mathbf{y},\mathbf{X},\mathbf{Z},\cdot$
\item $(\boldsymbol{\rho},\mathcal{I}_{\rho})\vert\mathbf{y},\mathbf{X},\mathbf{Z},\cdot$
\item $(\boldsymbol{\gamma},\mathcal{I}_{\gamma})\vert\mathbf{y},\mathbf{X},\mathbf{Z},\cdot$
\end{enumerate}
where $\cdot$ denotes all other model parameters.

\subsection{Prior distribution}

We assume prior independence between $\boldsymbol{\beta}$, $\boldsymbol{\gamma}$ and $\boldsymbol{\rho}$,
and let
\begin{align}
\boldsymbol{\beta} & \sim N\left(\boldsymbol{\mu}_{\beta},\Omega_{\beta}\right)\nonumber \\
\boldsymbol{\gamma} & \sim N\left(\boldsymbol{\mu}_{\gamma},\Omega_{\gamma}\right)\nonumber \\
\boldsymbol{\rho} & \sim N\left(\boldsymbol{\mu}_{\rho},\Omega_{\rho}\right),\label{eq:Prior1}
\end{align}
where $\Omega_{\beta}=\tau_{\beta}^{2}I_{p}$, $\Omega_{\gamma}=\tau_{\gamma}^{2}I_{q}$,
$\Omega_{\rho}=\tau_{\rho}^{2}\mathrm{Diag}\left(1,\frac{1}{2^{\zeta}},\frac{1}{3^{\zeta}}...,\frac{1}{k^{\zeta}}\right)$
and $\boldsymbol{\mu}_{\rho}=(r,0,...,0)^{T}$. The prior mean $\boldsymbol{\mu}_{\beta}$ is
set to 0 for all parameters, except for the term corresponding to
the intercept which is set to 800. The prior mean $\boldsymbol{\mu}_{\gamma}$
is set equal to 0 for all parameters. Note that the $N(\boldsymbol{\mu}_{\rho},\Omega_{\rho})$
prior centers the prior on the $AR(1)$ process $u_{t}=r\cdot u_{t-1}+\varepsilon_{t}$
, with coefficients corresponding to longer lags more tightly shrunk
toward zero. We also restrict the prior on $\boldsymbol{\rho}$ to the stationarity
region. The user is required to specify the prior hyperparameters
$\tau_{\beta},\tau_{\gamma},\tau_{\rho},r$ and $\zeta$. As default
values we use $\tau_{\beta}=\tau_{\gamma}=10$, $\tau_{\rho}=1$,
$r=0.5$ and $\zeta=1$, providing a rather uninformative prior. A
more complex prior, which for example allows for prior dependence
between $\boldsymbol{\beta}$ and $\boldsymbol{\gamma}$, can easily be incorporated into our
framework. 

\subsection{Variable selection}

Our MCMC algorithm presented in Section \ref{subsec:MCMC_HNGLM} performs
Bayesian variable selection among both sets of covariates, $\mathbf{x}_{t}$
(mean) and $\mathbf{z}_{t}$ (variance), {\color{black} using a spike and slab prior~\citep{george,kohn}}. We also use Bayesian variable selection in the AR noise process, thereby automatically learning
about the order $k$ of the AR process. The first element of $\boldsymbol{\beta}$
and $\boldsymbol{\gamma}$ (i.e., the intercepts in the mean and log variance,
respectively) are not subject to variable selection. To describe the
variable selection prior, let us focus on $\boldsymbol{\beta}$. Let $\boldsymbol{\beta}_{\mathcal{I_{\beta}}}$
denote the subset of regression coefficients selected by $\mathcal{I_{\beta}}$.
To allow for variable selection we take the prior for the unrestricted
$\boldsymbol{\beta}\sim N(\boldsymbol{\mu}_{\beta},\Omega_{\beta})$ and condition on the zeros
dictated by $\mathcal{I}_{\beta}$. Since all our prior covariance
matrices are diagonal, the conditional distributions are simply the
marginal distributions, e.g. $\boldsymbol{\beta}_{\mathcal{I_{\beta}}}\sim N\left(\boldsymbol{\mu}_{\beta,\mathcal{I_{\beta}}},\tau_{\beta}^{2}I_{p_{\mathcal{I_{\beta}}}}\right)$,
where $\boldsymbol{\mu}_{\beta,\mathcal{I_{\beta}}}$ is the subset of elements
of $\boldsymbol{\mu}$ corresponding to $\mathcal{I}_{\beta}$, and $p_{\mathcal{I_{\beta}}}$
is the number of elements in $\boldsymbol{\beta}_{\mathcal{I_{\beta}}}$. To complete
the variable selection prior we let the elements of $\mathcal{I_{\beta}}$
be apriori independent and Bernoulli distributed with $\mathrm{Pr}\left(I_{\beta,j}=1\right)=\pi_{\beta}$.
 {\color{black} The default values for $\pi_{\beta}$ and $\pi_{\gamma}$ are 0.5. The default value for $\pi_{\rho}$ is $0.5/\sqrt{k}$ for lag $k$, giving 0.5, 0.35, 0.29 and 0.25 for an AR(4) process. We also experiment with a hierarchical prior where the $\pi$ are assigned Beta priors, see below. The extension to a spatial prior on the variable selection indicators is also discussed below.}

\subsection{Variable selection in linear regression using MCMC\label{subsec:VariableSelectionLinearModel}}

This section describes how to simulate from the joint posterior of
the regression coefficients, and their variable selection indicators
in the Gaussian linear regression model with unit noise variance
\[
\mathbf{y}=\mathbf{X} \boldsymbol{\beta}+ \boldsymbol{\varepsilon},
\]
where $\boldsymbol{\varepsilon}=(\varepsilon_{1},...,\varepsilon_{T})^{T}$ and
$\varepsilon_{i}\overset{iid}{\sim}N(0,1)$. This will be an important
building block in our Metropolis-within-Gibbs algorithms described
in Sections \ref{eq:HIGLMfullSample} and \ref{subsec:MCMC_HNGLM}.
Similar to \citet{smithKohn1996}, we sample $\boldsymbol{\beta}$ jointly with
its variable selection indicators $\mathcal{I}$ (we drop the subscript
$\boldsymbol{\beta}$ here) by first generating from the marginal posterior $p(\mathcal{I}\vert\mathbf{y},\mathbf{X})$
followed by a draw from $p(\boldsymbol{\beta}\vert\mathcal{I},\mathbf{y},\mathbf{X})$.
A draw from $p(\boldsymbol{\beta}\vert\mathcal{I},\mathbf{y},\mathbf{X})$ is easily
obtained by sampling the non-zero elements of $\boldsymbol{\beta}$ as 
\begin{equation}
\boldsymbol{\beta}_{\mathcal{I}}\vert\mathcal{I},\mathbf{y},\mathbf{X}\sim N\left(\tilde{\boldsymbol{\beta}}_{\mathcal{I}},\left(\mathbf{X}_{\mathcal{I}}^{\prime}\mathbf{X}_{\mathcal{I}}+\Omega_{\mathcal{I}}^{-1}\right)^{-1}\right),\label{eq:PostBetaGivenI}
\end{equation}
where $\mathbf{X}_{\mathcal{I}}$ is the $T\times p_{\mathcal{I}}$ matrix
with covariates from the subset $\mathcal{I}$, {\color{black}$ \Omega_{\mathcal{I}} $ is the prior covariance for $\boldsymbol{\beta}_{\mathcal{I}} | \mathcal{I} $} and $\tilde{\boldsymbol{\beta}}_{\mathcal{I}}$
is given in Appendix B. A closed form expression for $p(\mathcal{I}\vert\mathbf{y},\mathbf{X})$,
the marginal posterior of $\mbox{\ensuremath{\mathcal{I}}}$, is given
in Appendix B, from which we can obtain $p(\mathcal{I}_{j}\vert\mathbf{y},\mathbf{X},\mathcal{I}_{-j})\propto p(\mathcal{I}\vert\mathbf{y},\mathbf{X})$,
where $\mathcal{I}_{-j}$ denotes $\mathcal{I}$ with the $j$th element
excluded. Simulating from the joint posterior of $\boldsymbol{\beta}$ and $\mathcal{I}$
is therefore acheived by simulating from each $p(\mathcal{I}_{j}\vert\mathbf{y},\mathbf{X},\mathcal{I}_{-j})$
in turn, followed by sampling of $\boldsymbol{\beta}_{\mathcal{I}}$ from \eqref{eq:PostBetaGivenI}.

\subsection{MCMC for the GLMH model\label{subsec:MCMC_HNGLM}}

\subsubsection*{Updating $(\boldsymbol{\beta},\mathcal{I}_{\beta})$}

To sample from the full conditional posterior of $(\boldsymbol{\beta},\mathcal{I}_{\beta})$
conditional on $\boldsymbol{\rho}$ and $\boldsymbol{\gamma}$, let us re-formulate the model
as
\begin{equation}
\tilde{y}_{t}=\tilde{\mathbf{x}}_{t}^{T}\boldsymbol{\beta}+\varepsilon_{t},\label{eq:ytildeRegression}
\end{equation}
where $\tilde{y}_{t}=\exp(-\mathbf{z}_{t}^{T}\boldsymbol{\gamma}/2)\rho(L)y_{t}$,
$\tilde{\mathbf{x}}_{t}=\exp(-\mathbf{z}_{t}^{T}\boldsymbol{\gamma}/2)\rho(L)\mathbf{x}_{t}$,
and $\rho(L)=1-\rho_{1}L-...-\rho_{k}L^{k}$ is the usual lag polynomial
in the lag operator $L^{k}y_{t}=y_{t-k}$ from time series analysis.
The Jacobian of the transformation $\mathbf{y}\rightarrow\tilde{\mathbf{y}}$
is $J\left(\mathbf{y}\rightarrow\tilde{\mathbf{y}}\right)=\exp\left(\boldsymbol{\gamma}^{T}\sum_{t=1}^{T}\mathbf{z}_{t}/2\right)$,
which can be seen as follows. The inverse transformation is $y_{t}=\rho^{-1}(L)\exp(\mathbf{z}_{t}^{T}\boldsymbol{\gamma}/2)\tilde{y}_{t}$,
where $\rho^{-1}(L)=1+\psi_{1}L+\psi_{2}L^{2}+...$ is the inverse
lag polynomial for some coeffcients $\psi_{1},\psi_{2},...$. This
system of equations is recursive so the Jacobian is $\left|\prod_{t=1}^{T}\frac{\partial y_{t}}{\partial\tilde{y}_{t}}\right|$
and $\frac{\partial y_{t}}{\partial\tilde{y}_{t}}=\exp(\mathbf{z}_{t}^{T}\boldsymbol{\gamma}/2)$
which proves the result. Note that $J(\mathbf{y}\rightarrow\tilde{\mathbf{y}})$
does not depend on $\boldsymbol{\beta}$ and can therefore be ignored when deriving
the full conditional posterior of $\boldsymbol{\beta}$. Now, $\boldsymbol{\beta}$ in \eqref{eq:ytildeRegression}
are the coefficients in a linear regression with unit noise variance
and we can draw from the full conditional $p(\boldsymbol{\beta},\mathcal{I}_{\beta}\vert\mathbf{y},\mathbf{X},\mathbf{Z},\cdot)$
as described in Section \ref{subsec:VariableSelectionLinearModel}
with $\mathbf{y}$ and $\mathbf{X}$ replaced by $\tilde{\mathbf{y}}$
and $\mathbf{\tilde{X}}$, respectively.

\subsubsection*{Updating $(\rho,\mathcal{I}_{\rho})$}

The AR process can be rewritten as
\[
\tilde{u}_{t}=\tilde{U}_{t}\boldsymbol{\rho}+\varepsilon_{t},
\]
where $\tilde{u}_{t}=\exp(-\mathbf{z}_{t}^{T}\boldsymbol{\gamma}/2)u_{t}$. The
Jacobian of this transformation is $J\left(\mathbf{u}\rightarrow\tilde{\mathbf{u}}\right)=\exp\left(\boldsymbol{\gamma}^{T}\sum_{t=1}^{T}\mathbf{z}_{t}/2\right)$ which does not depend on $\boldsymbol{\rho}$ and can therefore be ignored when
updating $\boldsymbol{\rho}$. Now, $\boldsymbol{\rho}$ are the coefficients in a linear regression
with unit noise variance and we can draw from the full conditional
$p(\boldsymbol{\rho},\mathcal{I}_{\rho}\vert\mathbf{y},\mathbf{X},\mathbf{Z},\cdot)$
as described in Section \eqref{subsec:VariableSelectionLinearModel}.

\subsubsection*{Updating $(\boldsymbol{\gamma},\mathcal{I}_{\rho})$}

The full conditional posterior of $\left(\boldsymbol{\gamma},\mathcal{I}_{\gamma}\right)$
is a complicated distribution which we can not easily sample from.
However, it is clear from the model
\[
u_{t}=\rho_{1}u_{t-1}+...+\rho_{k}u_{t-k}+\exp(\mathbf{z}_{t}^{T}\boldsymbol{\gamma}/2)\cdot\varepsilon_{t},
\]
that the conditional likelihood of $\boldsymbol{\gamma}$ is of the form described
in \citet{Villani2012GSM} where the observations (the $u_{t}$ in
this case) are conditionally independent and $\boldsymbol{\gamma}$ enters each
factor in the likelihood linearly ($\mathbf{z}_{t}^{T}\boldsymbol{\gamma}$) through
a scalar valued quantity $\varphi_{t}=\exp(\mathbf{z}_{t}^{T}\boldsymbol{\gamma}/2)$.
The MCMC update with a finite step Newton proposal with variable selection
described in \citep{villani2009sagm,Villani2012GSM} can therefore
be used. In fact, \citet{villani2009sagm} contains the details for
the Gaussian heteroscedastic regression, which is exactly the model
when we condition on $\boldsymbol{\beta}$ (since $\mathbf{u}$ is then known).
The algorithm in \citep{Villani2012GSM} proposes $\boldsymbol{\gamma}$ and $\mathcal{I}_{\gamma}$
jointly by randomly changing a subset of the indicators in $\mathcal{I}_{\gamma}$
followed by a proposal from $\boldsymbol{\gamma}\vert\mathcal{I}_{\gamma}$ using
a multivariate-\emph{t} distribution tailored to the full conditional
posterior. The tailoring is acheived by taking a small number of Newton
steps toward the posterior mode, and using the negative inverse Hessian
at the terminal point as the covariance matrix in the multivariate-\emph{t}
proposal distribution. The update is fast, since the Jacobian and
Hessian can be computed in closed form using the chain rule and compact
matrix algebra. It is also possible to compute the expected Hessian
(Fisher information) in closed form. The expected Hessian tends to
be more stable numerically with only marginally worse tailoring to
the posterior. Note also that the Newton iterations always start from
the current value of $\boldsymbol{\gamma}$, which is typically not far from the
mode, so even one or two Newton steps are usually sufficient. We refer
to \citet{Villani2012GSM} for details of the general algorithm, and
to \citet{villani2009sagm} for expressions of the Jacobian, Hessian
and expected Hessian for $\boldsymbol{\gamma}$. 

{\color{black} 

\subsubsection*{Updating $\pi_{\beta}$ and  $\pi_{\gamma}$}

The inclusion probabilities $\pi_{\beta}$ and  $\pi_{\gamma}$ for the variable selection can also be updated in every MCMC iteration\citep{kohn} (updating $\pi_{\rho}$ is in principle straightforward, but there is very little information about $\pi_{\rho}$, due to the low number of AR parameters). Let the prior for $\pi_{\beta}$ and $\pi_{\gamma}$ be $\mathrm{Beta}(a,b)$. The conditional posterior for $\pi_{\beta}$ is then given by $\mathrm{Beta}(a + \sum_{j=1}^p \mathcal{I}_{\beta,j},b + p - \sum_{j=1}^p \mathcal{I}_{\beta,j})$, where $p$ is the number of covariates and $\mathcal{I}_{\beta,j}$ is the binary indicator varible for covariate $j$. The posterior for $\pi_{\gamma}$ is defined analogously. We use $a = b = 3$ which gives a prior with a mean of 0.5. The complete algorithm becomes
\begin{enumerate}
\item $(\boldsymbol{\beta},\mathcal{I}_{\beta})\vert\mathbf{y},\mathbf{X},\mathbf{Z},\cdot$
\item $(\pi_{\beta})\vert\mathbf{y},\mathbf{X},\mathbf{Z},\cdot$
\item $(\boldsymbol{\rho},\mathcal{I}_{\rho})\vert\mathbf{y},\mathbf{X},\mathbf{Z},\cdot$
\item $(\boldsymbol{\gamma},\mathcal{I}_{\gamma})\vert\mathbf{y},\mathbf{X},\mathbf{Z},\cdot$
\item $(\pi_{\gamma})\vert\mathbf{y},\mathbf{X},\mathbf{Z},\cdot$
\end{enumerate}
where $\cdot$ denotes all other model parameters.

\subsubsection*{Spatial variable selection prior}

Since $\boldsymbol{\beta}$ and $\boldsymbol{\rho}$ both appear as coefficients in linear regressions (conditional on the other parameters), it is straightforward to extend our variable selection for $\mathcal{I}_{\beta}$ and $\mathcal{I}_{\rho}$ to have a spatial binary Markov random field prior following~\citet{fahrmeir}, but would naturally add to the processing time. A spatial prior on $\mathcal{I}_{\gamma}$ is more difficult since $\boldsymbol{\gamma}$ does not appear linearly in the model, even conditional on the other parameters, and can therefore not be integrated out analytically as in~\citet{fahrmeir}. We leave such an extension to future work.

}

\section{Implementation}

A drawback of using MCMC is that processing of a single fMRI dataset
can take several hours \citep{Woolrich2004,siden}. Our implementation
of the heteroscedastic GLM is therefore written in C++, using the
Eigen library \citep{eigenweb} for all matrix operations. The random
number generators available in the C++ standard library (available
from C++ 2011) were used, together with the Eigen library, to make
random draws from multivariate distributions. The OpenMP (Open Multi
Processing) library was used to take advantage of all the CPU cores,
by analyzing several voxels in parallel. For all analyses the number
of Newton steps is set to 2. To lower the processing time, the variable
selection indicators for the variance covariates are only updated
in 60\% of the draws. {\color{black} See Appendix A for more information about the implementation. The code is available at https://github.com/wanderine/HeteroscedasticfMRI }

\section{Results}

\subsection{Simulated data}

\subsubsection{GLMH vs Bayesian GLM with homoscedastic noise}

To verify that the heteroscedastic model works as expected, and to
compare it to a homoscedastic model for data with a known activity
pattern, the algorithms were applied to simulated data with homoscedastic
and heteroscedastic noise. The simulated data were created using (posterior
mean) beta estimates from {\color{black} spatially smoothed} real fMRI data (with several motion spikes),
together with the applied design matrix, to create a timeseries in
each voxel. The design matrix consisted of an intercept, time trends
for modeling drift (linear, quadratic and cubic), activity covariates,
estimated head motion parameters and their temporal derivative (in
total 16 covariates in addition to the activity covariates{\color{black}, see Figure~\ref{fig:XZexample} for an example}). {\color{black} The simulated data thereby contain spatial correlation as well as correlation between the covariates.} Beta values for active voxels were generated from a $abs(N(0,9))+3$ distribution,
and for non-active voxels from a $N(0,0.06)$ distribution. The simulated
activity is thereby very easy to detect, and the difficult part is
to model the heteroscedastic noise. 

\begin{figure}
\includegraphics[scale=0.3]{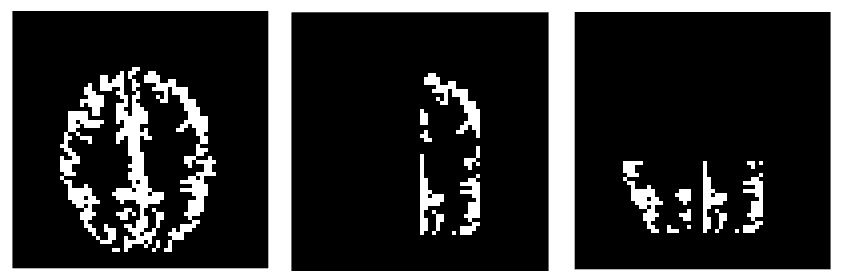}

\caption{Left: A mask for gray matter voxels. Middle: Voxels with simulated
activity. Right: Voxels with heteroscedastic noise. The simulated
data consists of four regions; active voxels with homoscedastic or
heteroscedastic noise, and non-active voxels with homoscedastic or
heteroscedastic noise. \label{fig:masks}}
\end{figure}

\begin{figure}
\includegraphics[scale=0.45]{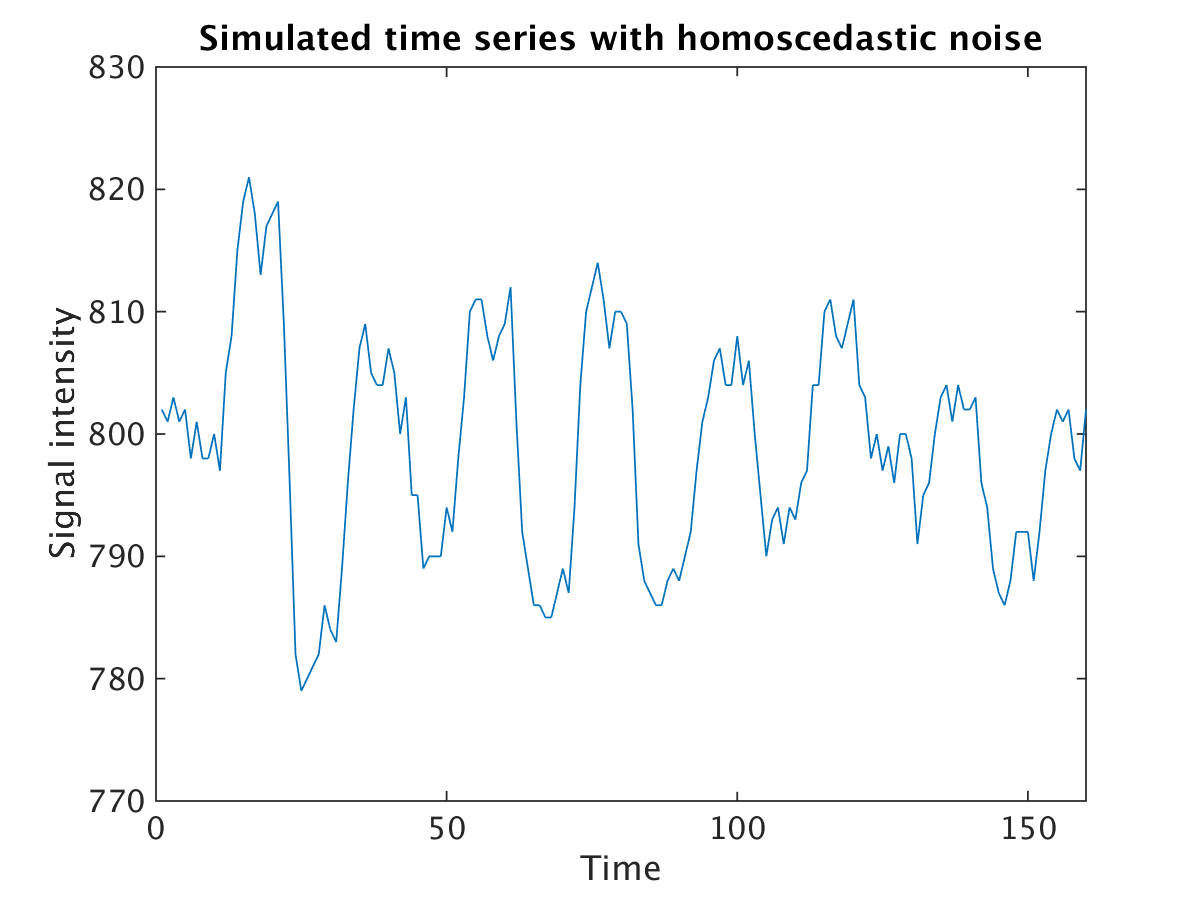}
\includegraphics[scale=0.45]{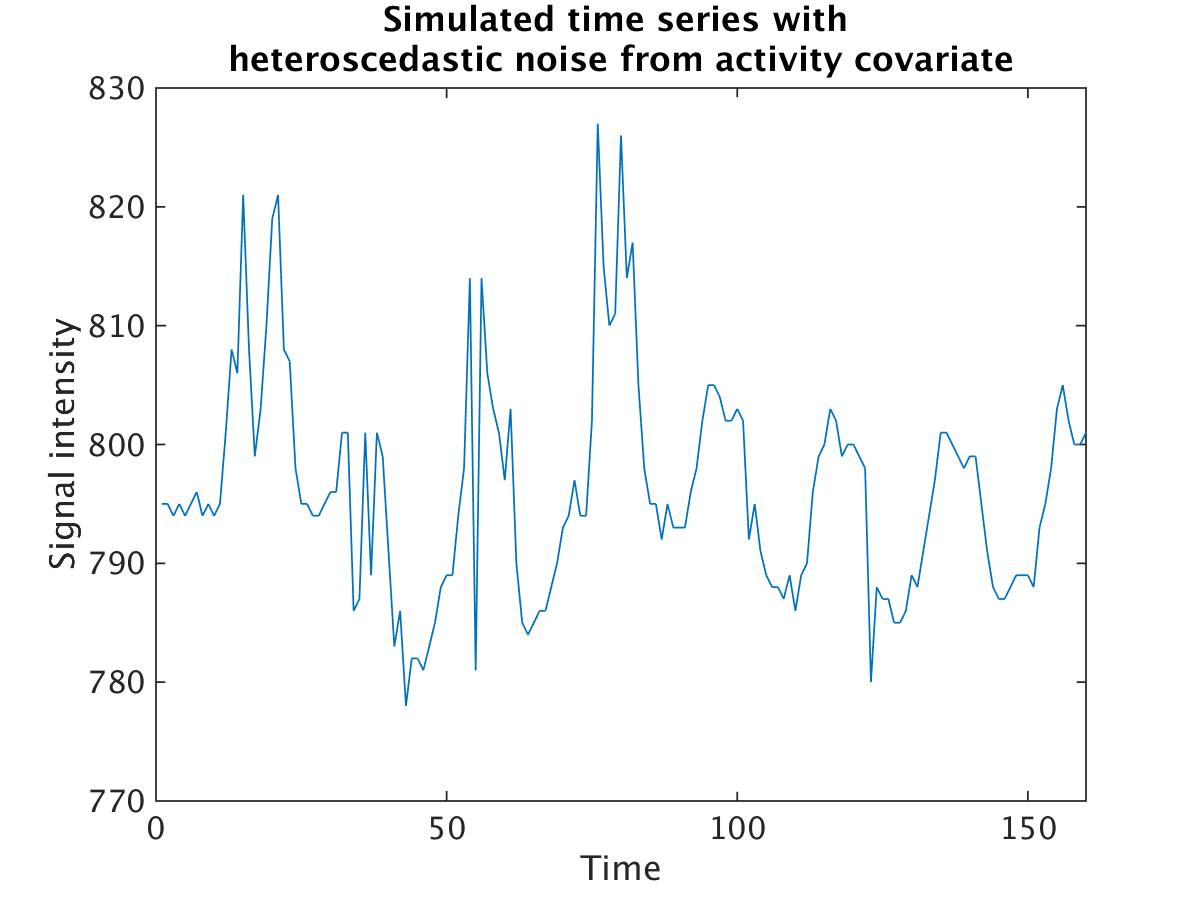}
\includegraphics[scale=0.45]{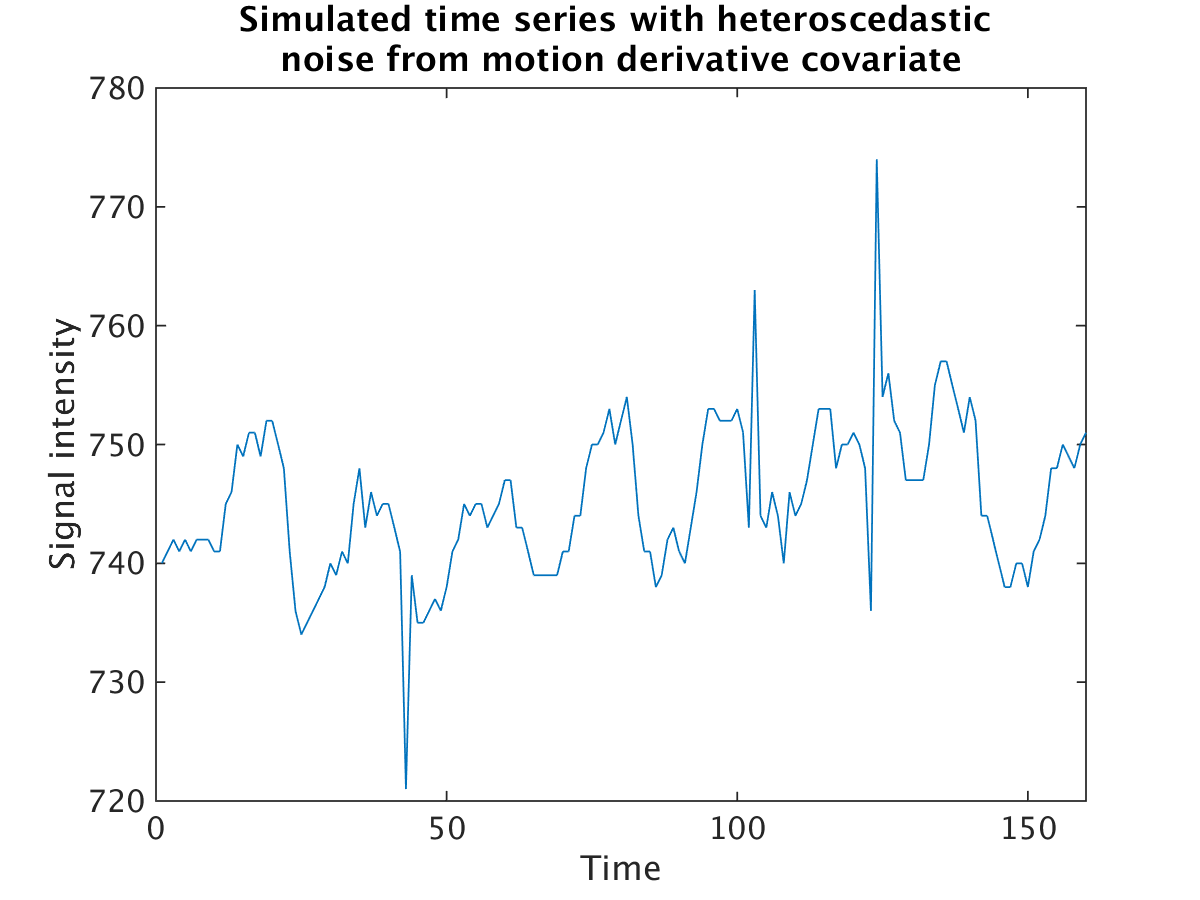}
\caption{{\color{black} Top: A simulated time series with homoscedastic noise. Middle: A simulated time series with heteroscedastic noise, where the variance is modelled as a function of an activity covariate (the variance is higher for the first part of the dataset, which corresponds to the first activity covariate). Bottom: A simulated time series with heteroscedastic noise, where the variance is modelled as a function of a covariate representing the temporal derivative of the head motion. In all cases the simulated brain activity is rather strong, but the heteroscedastic noise makes it difficult to detect activity using homoscedastic methods. } \label{fig:simulatedtimeseries}}
\end{figure}

For approximately half of the active voxels, heteroscedastic noise
was added according to Equation 1. For one covariate at a time (either
an activation or head motion covariate), the corresponding $\gamma$
parameter was set to 1, 2, or 3. For one covariate representing the
(absolute value of the) temporal derivative of the head motion, the
$\gamma$ parameter was instead set to 1, 1.25 or 1.5 (as motion spikes
can be rather large, and thereby make the simulation unrealistic).
To simulate simultaneous heteroscedasticity from several covariates,
the $\gamma$ parameters for the activity and the head motion covariates
were simultaneously set to 1, 2, or 3, while the $\gamma$ parameter
for the derivated head motion covariate was set to 1.25 for all cases.
The $\gamma$ parameter for the intercept covariate was always set
to 1, and all other $\gamma$ parameters were set to 0. For all other
voxels, homoscedastic noise was added ($\gamma$ = 1 for the intercept
only). The four autocorrelation parameters were set to 0.4, 0.2, 0.1
and 0.05, respectively. The simulated data thereby consists of four
regions; active voxels with homoscedastic or heteroscedastic noise,
and non-active voxels with homoscedastic or heteroscedastic noise.
To lower the processing time, only a single slice of data was simulated.
See Figure~\ref{fig:masks} for the gray matter mask, the mask for
active voxels and the mask for voxels with heteroscedastic noise. 
{\color{black} Figure~\ref{fig:simulatedtimeseries} shows one simulated time series with homoscedastic noise, and two simulated time series with heteroscedastic noise.}

For each simulated dataset, the analysis was performed (i) including
only an intercept for the variance (i.e., a homoscedastic model) and
(ii) including all covariates for the variance (i.e., a heteroscedastic
model). In both cases all covariates (except the intercept) were standardized,
to have zero mean and variance 1. For the mean covariates, the original
temporal derivative of the head motion parameters was used. For the
variance covariates, the absolute value was used instead, as the variance
should always increase at a motion spike regardless of the direction
(positive or negative){\color{black}, see Figure~\ref{fig:XZexample} for an example of the covariates for the mean and the variance}. For both models, a fourth order AR model was
used in each voxel. Variable selection was performed on all covariates
(mean and variance), except for the intercept, as well as for the
four AR parameters. Stationarity was enforced for the AR parameters,
by discarding draws where the absolute value of any eigenvalue of
the companion matrix is larger than or equal to 1. For each voxel,
a total of 1,000 draws were used for MCMC burn-in and another 1,000
draws were saved for inference.

\begin{figure}
\includegraphics[scale=0.45]{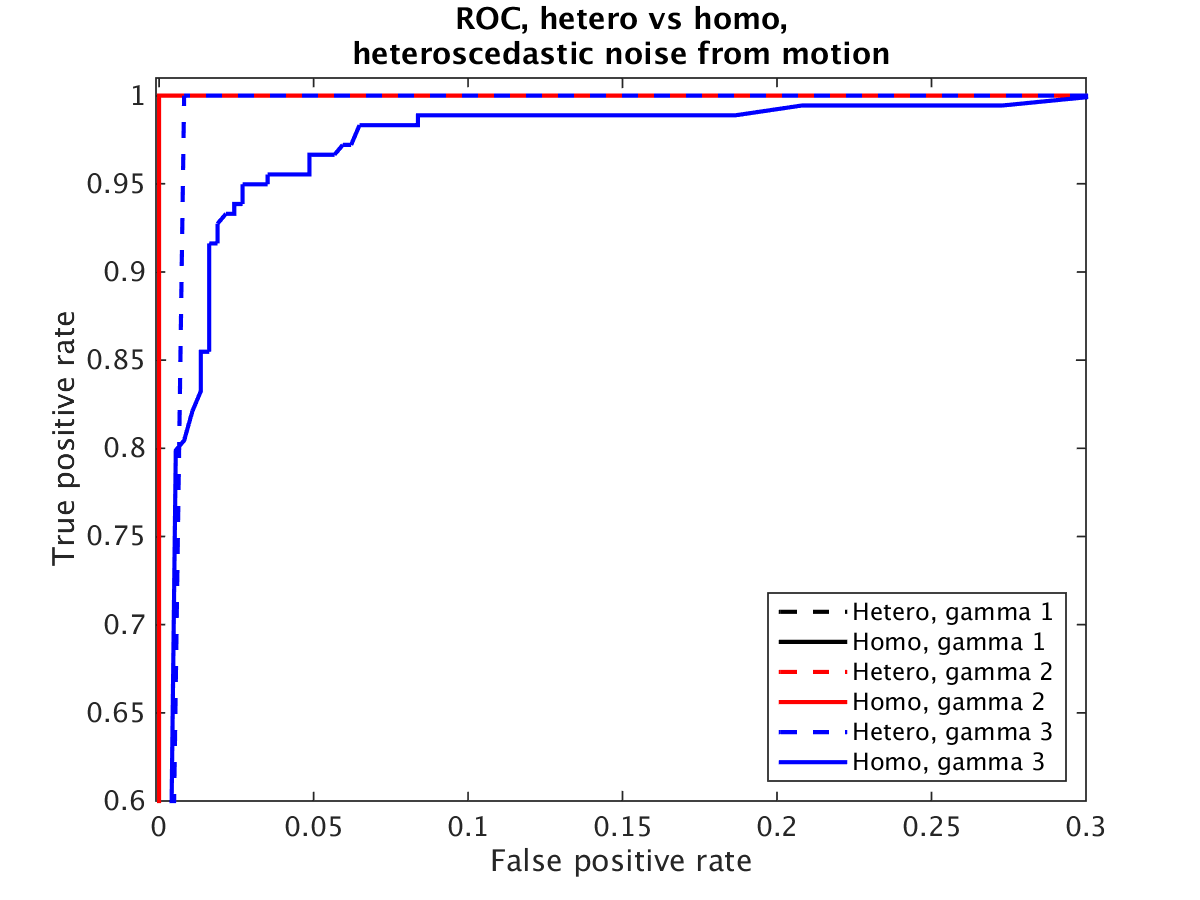}

\caption{ROC curves for simulated data with heteroscedastic noise from one
motion covariate, and different levels of heteroscedasticity. Both
models perform well for low levels of heteroscedasticity, but the
homoscedastic model performs worse for high levels of heteroscedasticity.
\label{fig:simulationresults1}}
\end{figure}

\begin{figure}
\includegraphics[scale=0.45]{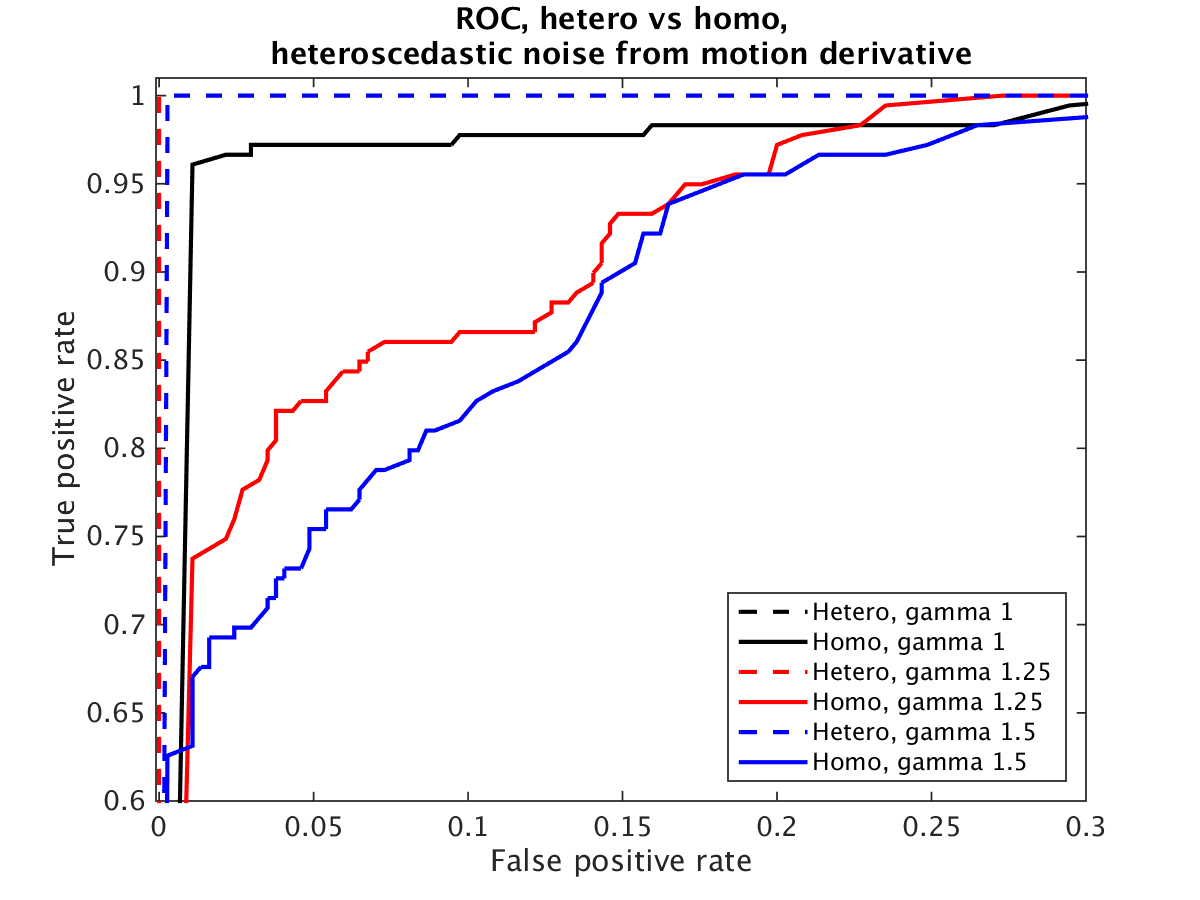}

\caption{ROC curves for simulated data with heteroscedastic noise from the
temporal derivative of one motion covariate, and different levels
of heteroscedasticity. The homoscedastic model has a lower performance,
and fails to detect a large portion of the active voxels.\label{fig:simulationresults2}}
\end{figure}

\begin{figure}
\includegraphics[scale=0.45]{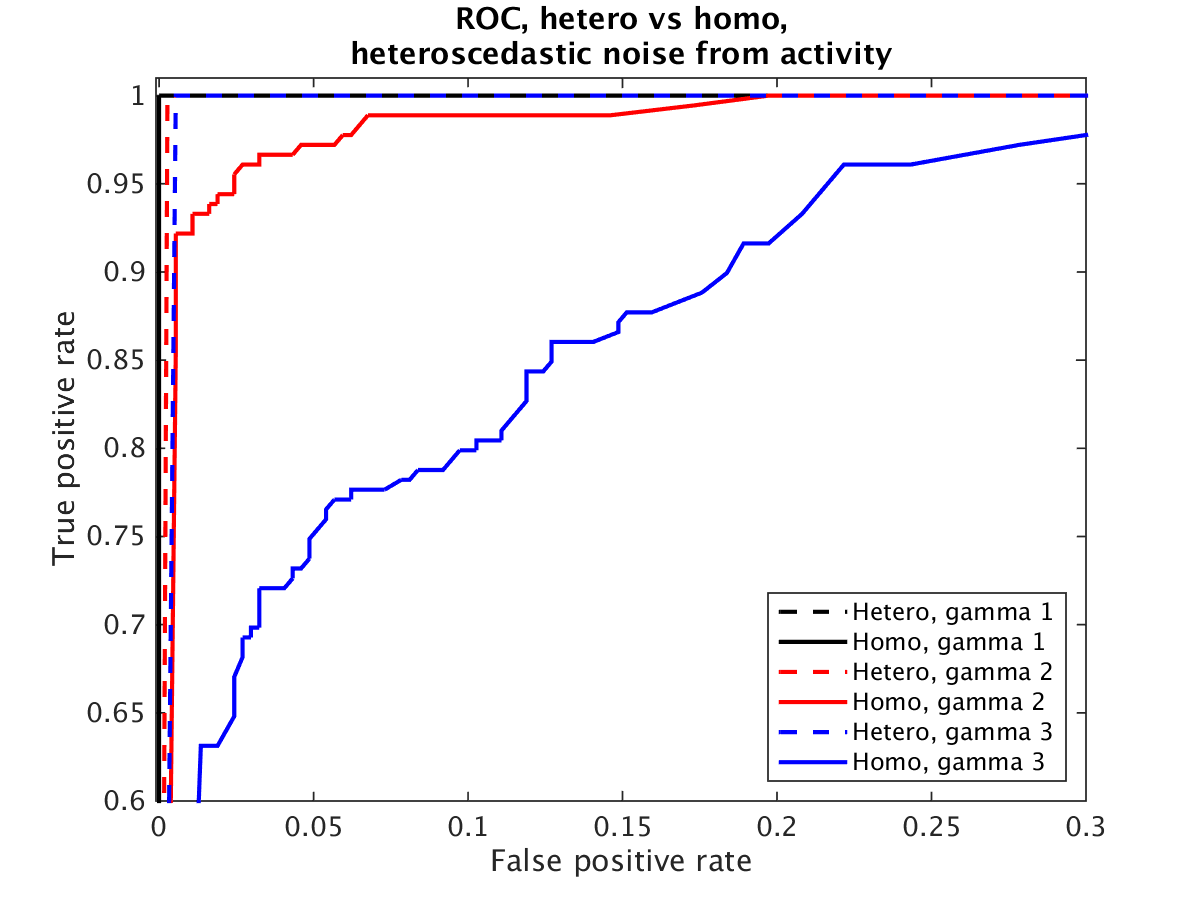}

\caption{ROC curves for simulated data with heteroscedastic noise from one
activity covariate, and different levels of heteroscedasticity. The
homoscedastic model has a lower performance, and fails to detect a
large portion of the active voxels.\label{fig:simulationresults3}}
\end{figure}

\begin{figure}
\includegraphics[scale=0.45]{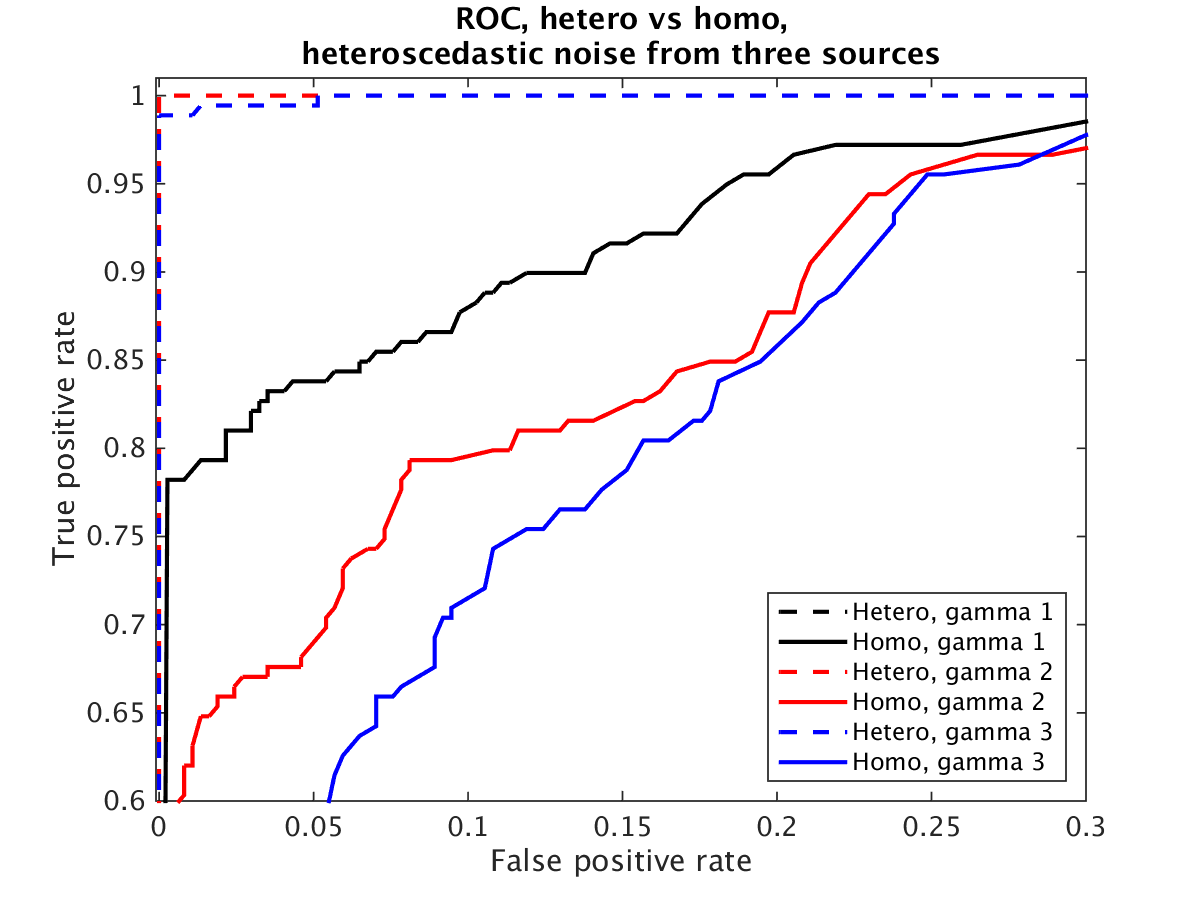}

\caption{ROC curves for simulated data with heteroscedastic noise from three
simultaneous sources (motion, motion derivative, activity), and different
levels of heteroscedasticity. The homoscedastic model has a much lower
performance, and fails to detect a large portion of the active voxels.\label{fig:simulationresults4}}
\end{figure}

Figures~\ref{fig:simulationresults1} -~\ref{fig:simulationresults4}
show receiver operating characteristic (ROC) curves for the two models,
for different types (activity, motion, motion derivative, all) and
levels ($\gamma$ = 1, 2 or 3) of heteroscedasticity. The ROC curves
were generated by varying the threshold for the posterior probability
maps (PPMs) from 0.01 to 1.00. It is clear that both models detect virtually all the active voxels for low levels of heteroscedasticity, while the homoscedastic model fails to detect a large portion of the
active voxels with heteroscedastic noise for higher levels of heteroscedasticity.
The posterior inclusion probabilities for the variance parameters
($\gamma$) indicate that the heteroscedastic model in virtually all
voxels only includes the covariates that were used to generate the
heteroscedastic noise (not shown).

\subsubsection{GLMH vs weighted least squares }

\begin{figure}
\includegraphics[scale=0.45]{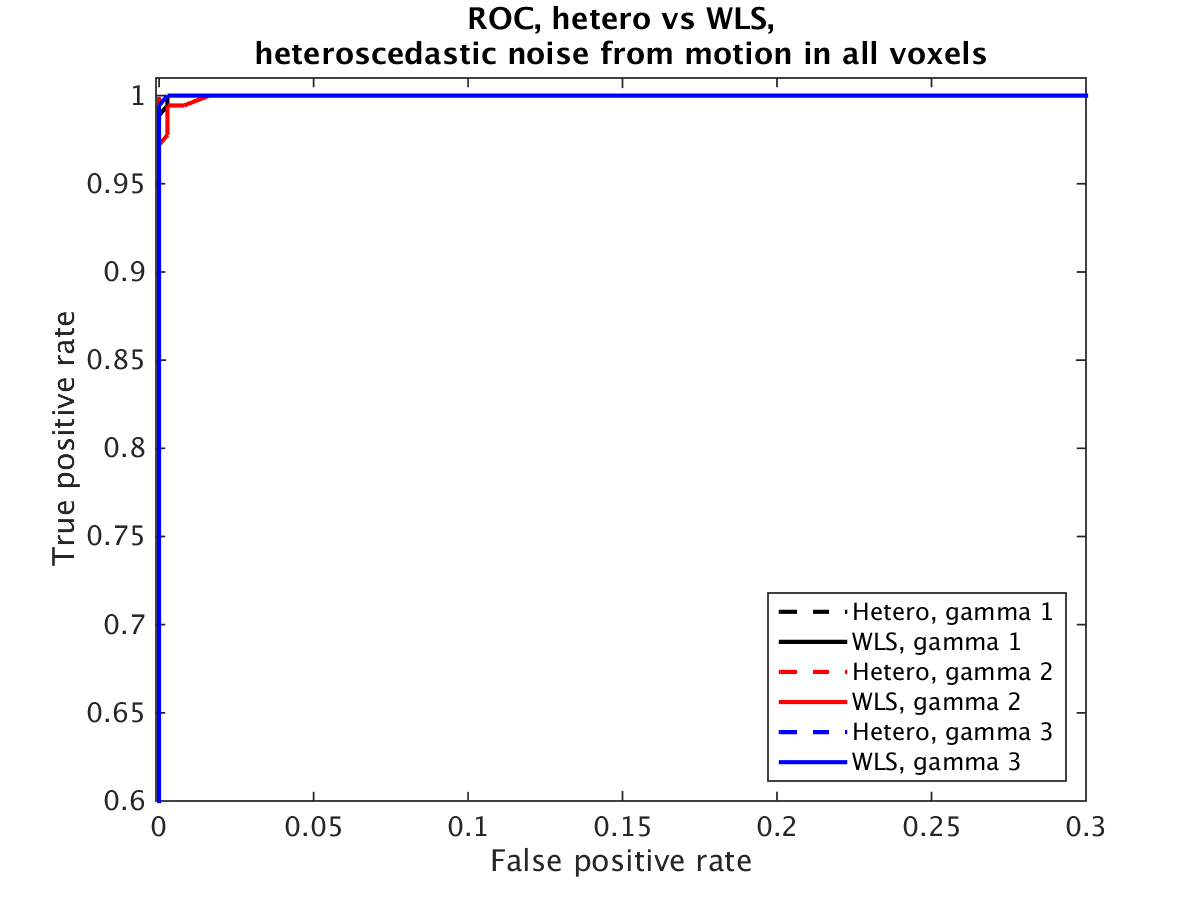}

\caption{ROC curves for simulated data with heteroscedastic noise in all voxels,
generated by one motion covariate. Both approaches perform well for
all levels of heteroscedasticity. \label{fig:simulationresults5}}
\end{figure}

\begin{figure}
\includegraphics[scale=0.45]{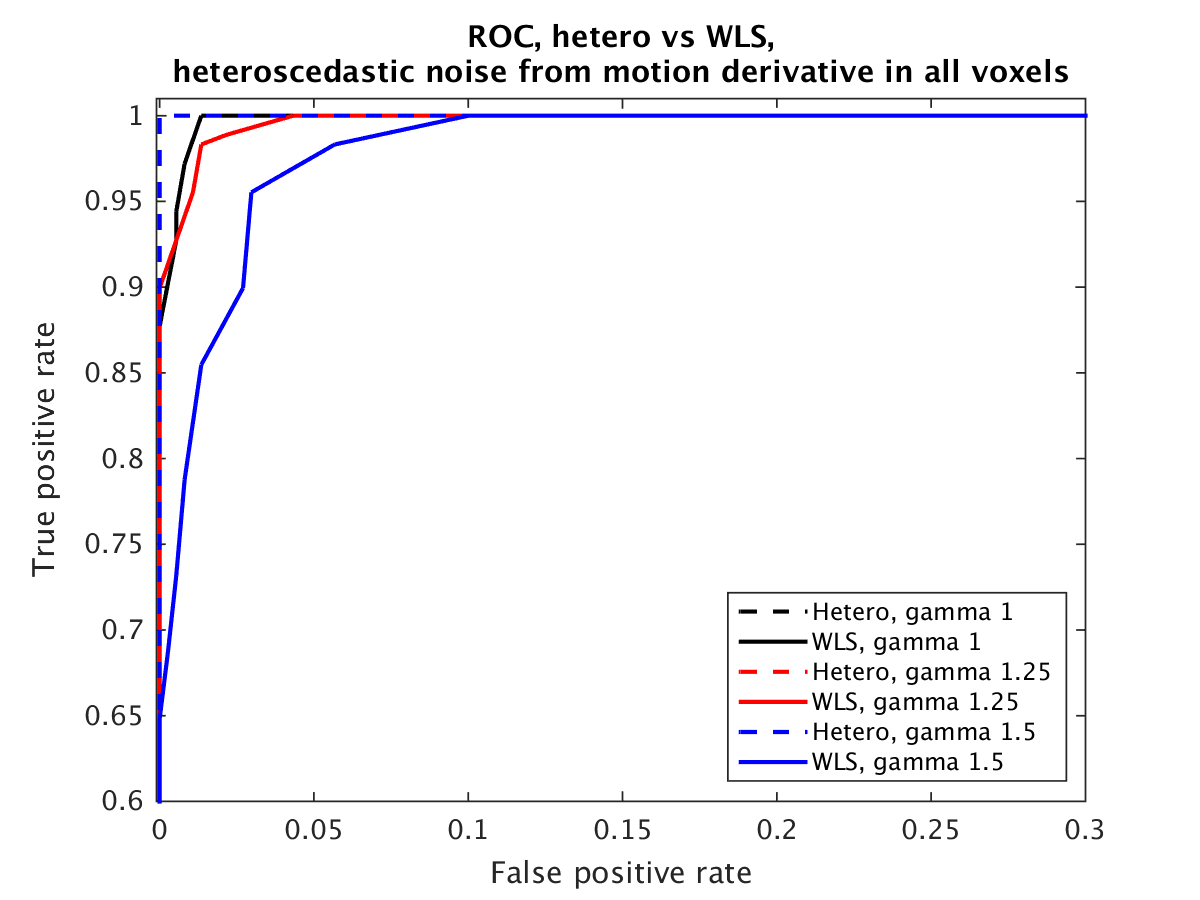}

\caption{ROC curves for simulated data with heteroscedastic noise in all voxels,
generated by the temporal derivative of one motion covariate. Both
approaches perform well, but the hetero approach works better for
higher levels of heteroscedasticity.\label{fig:simulationresults6}}
\end{figure}

\begin{figure}
\includegraphics[scale=0.45]{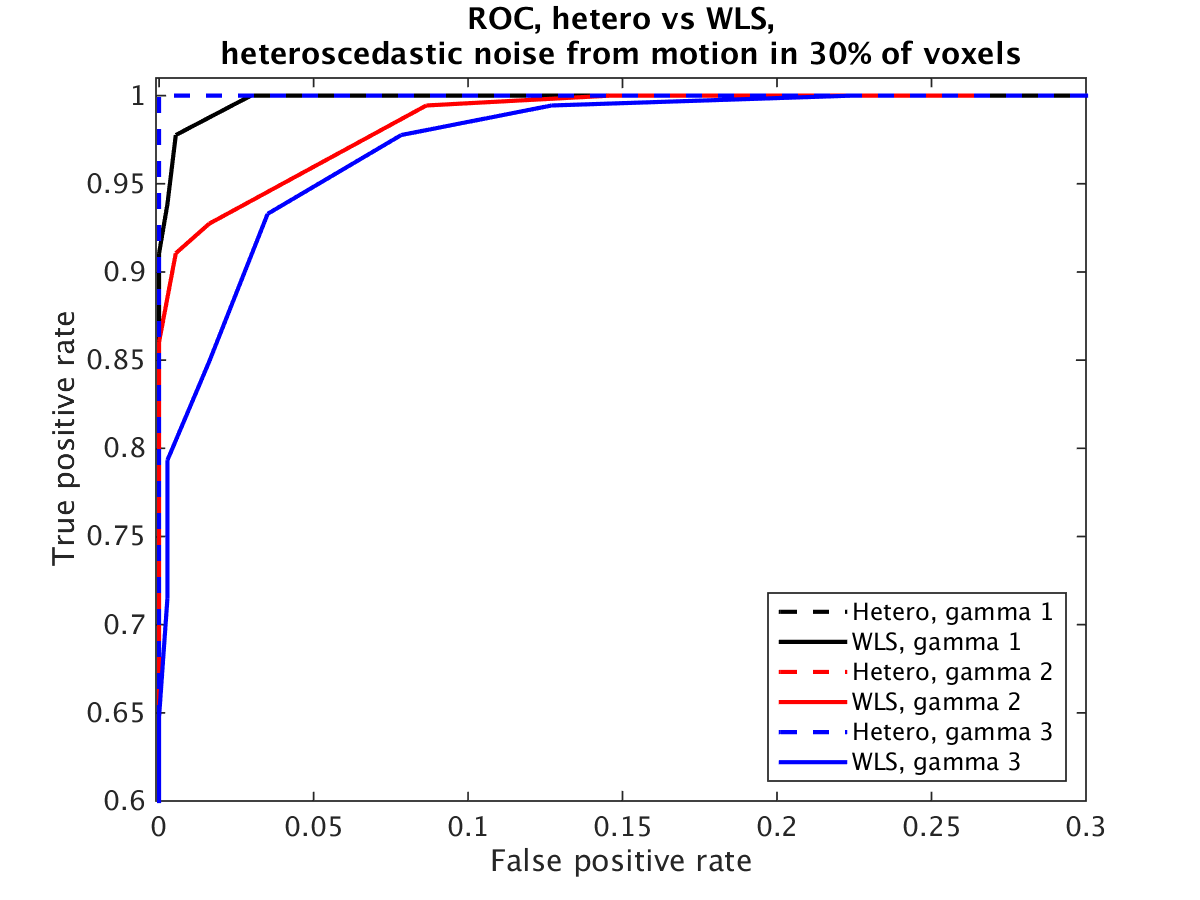}

\caption{ROC curves for simulated data with heteroscedastic noise in 30\% of
the voxels, generated by one motion covariate. Compared to heteroscedastic
noise in all voxels, the WLS approach has a slightly lower performance.
\label{fig:simulationresults7}}
\end{figure}

\begin{figure}
\includegraphics[scale=0.45]{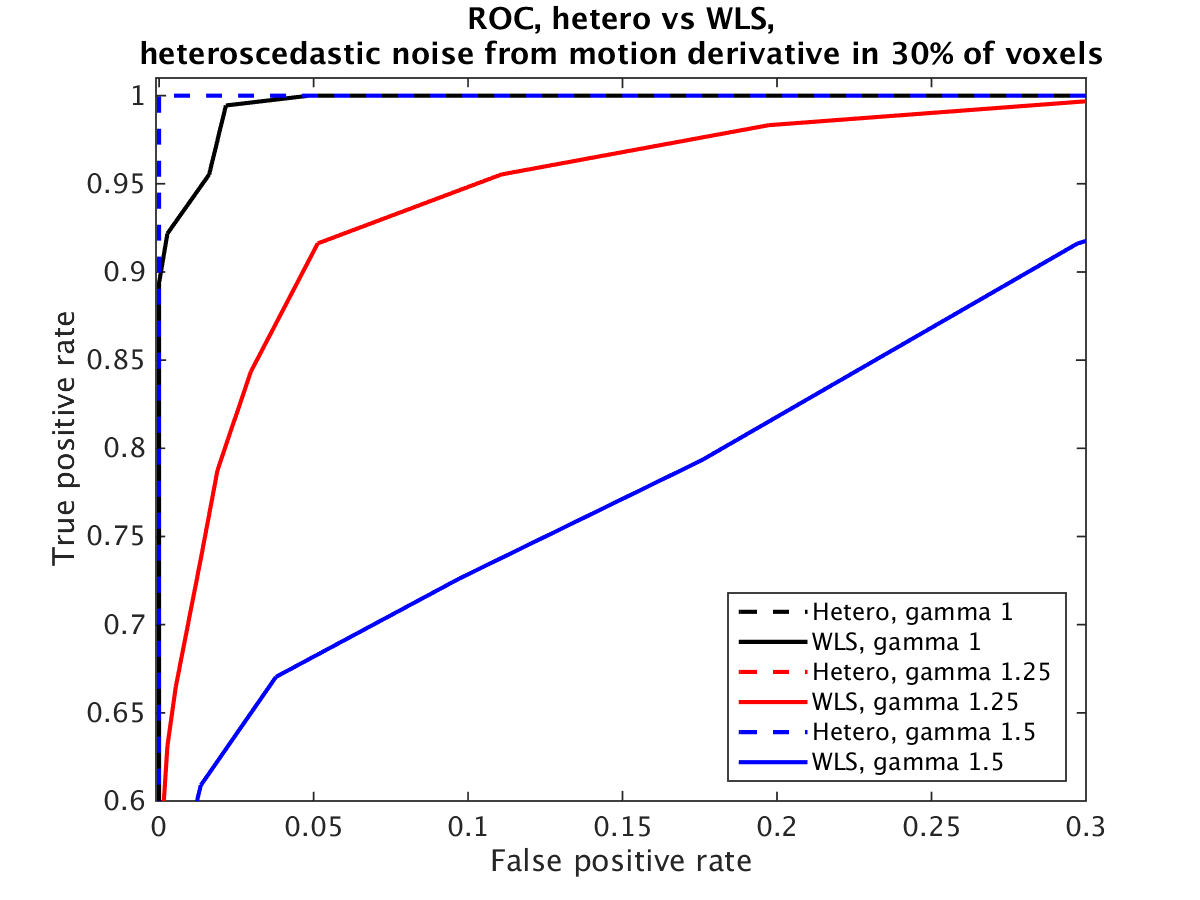}

\caption{ROC curves for simulated data with heteroscedastic noise in 30\% of
the voxels, generated by the temporal derivative of one motion covariate.
Compared to heteroscedastic noise in all voxels, the WLS approach
fails to detect a large portion of the active voxels.\label{fig:simulationresults8}}
\end{figure}

To compare the heteroscedastic model to the weighted least squares
(WLS) approach proposed by \citet{Diedrichsen2005}, where a single
weight is estimated for each volume, two additional datasets were
simulated (using the same activity mask as above). For the first dataset,
the same heteroscedastic noise was added to all voxels. For the second
dataset, heteroscedastic noise was added to only 30\% of the voxels
(using the same hetero mask as above). The simulation was performed
to generate different types (motion, motion derivative) and levels
($\gamma$ = 1, 2 or 3) of heteroscedasticity. As the two approaches
use different models for the temporal autocorrelation, the four AR
parameters were set to 0, to focus solely on the heteroscedasticity.
To mimic the analysis by \citet{Diedrichsen2005}, no motion regressors
were used in the design matrix for the WLS approach. Bayesian $t$-scores
(posterior mean / posterior standard deviation) were calculated for
the heteroscedastic model, and compared to regular $t$-scores from
the WLS approach. Figures~\ref{fig:simulationresults5} - ~\ref{fig:simulationresults8}
show ROC curves for the two approaches. Both approaches work well
when the same heteroscedastic noise is present in all voxels, but
the WLS approach fails to detect a large portion of the activity when
the heteroscedastic noise is only present in 30\% of the voxels.

\clearpage


\clearpage


\clearpage

\subsection{Application to real data}

Three datasets from the OpenfMRI project\citep{poldrack,poldrackNN}
were analyzed using both the homoscedastic and heteroscedastic noise
models. The datasets include experiments on rhyme judgment\footnote{https://openfmri.org/dataset/ds000003/},
living-nonliving judgment\footnote{https://openfmri.org/dataset/ds000006/}
and mixed gambles\footnote{https://openfmri.org/dataset/ds000005/}\citep{mixedgambles}.

In the rhyme judgment task, stimuli were presented in pairs (consisting
of either words or pseudo-words) and the subject was asked whether
the pair of stimuli rhymed with one another. The dataset consists
of 13 subjects and two different conditions: words and pseudo-words. 

In the living/nonliving judgment task, subjects were presented with
words in either plain or mirror-reversed format, and asked whether
the stimulus referred to a living or nonliving object. The data set
consists of 14 subjects and 4 different conditions: mirror-reversed
trials preceded by a plain text trial, mirror-reversed trials preceded
by a mirror-reversed trial, plain-text trials preceded by a mirror-reversed
trial, and plain-text trials preceded by a plain-text trial. A fifth
covariate is used to represent failed (junk) trials. 

Finally, in the mixed gambles task, subjects were presented with gambles
in which they have a 50\% chance of gaining and a 50\% chance of losing
money, where the potential gain and loss varied across trials. The
subject then decided whether or not to accept the gamble. The data
set consists of 16 subjects and 4 different conditions: task, parametric
gain, parametric loss, and distance from indifference point. For more
details on the 3 datasets we refer to the OpenfMRI website (https://openfmri.org).

{\setlength{\parindent}{0cm}

\subsubsection{Single subject analysis}

Prior to statistical analysis, the BROCCOLI software \citep{Eklund2014}
was used to perform motion correction and 6 mm FWHM smoothing. For
each subject, the analysis was performed as described for the simulated
data (16 covariates + activity covariates, for both mean and variance). {\color{black} For each dataset, the analysis was performed (i) including only an intercept for the variance (i.e., a homoscedastic model) and (ii) including all covariates for the variance (i.e., a heteroscedastic model).} Only gray matter voxels were analyzed to lower processing time. All results were finally transformed to MNI space, by combining T1-MNI
and fMRI-T1 transforms. 

Figure~\ref{fig:firstlevelresults} shows PPMs for one subject from
the rhyme judgment dataset and one subject from the mixed gambles
dataset; the heteroscedastic model tends to detect more brain activity
compared to the homoscedastic model. Figures~\ref{fig:ppmdiff1} -
~\ref{fig:ppmdiff3} summarize the number of voxels where the difference
between the heteroscedastic PPM and the homoscedastic PPM is larger
than 0.5, for the three different datasets. The largest PPM differences
are found in the rhyme judgment dataset, which contains the highest
number of motion spikes. Figure~\ref{fig:stdcomparison} shows a comparison
between the estimated homoscedastic and heteroscedastic standard deviation
for a single time series; the heteroscedastic standard deviation is
much higher for time points close to motion spikes, but lower for
time points with little head motion. The homoscedastic model struggles
to find a single variance to fit both time points with and without
motion, thereby ending up inflating the variance at times with little
or no motion. The heteroscedastic model can have a lower variance
at timeperiods with little motion, and is therefore able to detect
more brain activity. Figures~\ref{fig:varcovariates1} - ~\ref{fig:varcovariates3}
show the number of voxels, for each dataset, where the posterior inclusion
probability is larger than 90\% for the variance covariates. The temporal
derivative of the head motion parameters are clearly the most important
covariates for modeling the variance.

\begin{figure*}
\hspace{1.6cm}\begin{huge}Hetero\hspace{4.5cm}Homo\hspace{3.5cm}Hetero
- Homo\end{huge}\vspace{0.5cm}

\begin{minipage}[t]{0.3\textwidth}%
\includegraphics[scale=0.9]{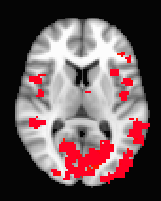}%
\end{minipage}\hfill{}%
\begin{minipage}[t]{0.3\textwidth}%
\includegraphics[scale=0.9]{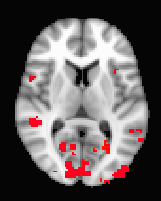}%
\end{minipage}\hfill{}%
\begin{minipage}[t]{0.3\textwidth}%
\includegraphics[scale=0.9]{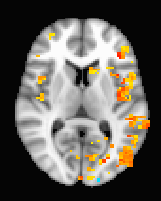}%
\end{minipage}\medskip{}
\hfill{}%
\begin{minipage}[t]{0.3\textwidth}%
\includegraphics[scale=0.8]{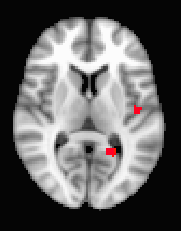}%
\end{minipage}\hfill{}%
\begin{minipage}[t]{0.3\textwidth}%
\includegraphics[scale=0.8]{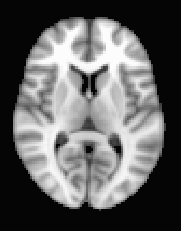}%
\end{minipage}\hfill{}%
\begin{minipage}[t]{0.3\textwidth}%
\includegraphics[scale=0.8]{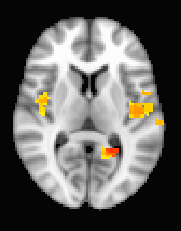}%
\end{minipage}\caption{Single subject posterior probability maps (PPMs) for the rhyme judgment
and mixed gambles datasets. From left to right: PPM for the heteroscedastic
model, PPM for the homoscedastic model, PPM hetero - PPM homo. The
hetero and the homo PPMs are thresholded at Pr = 0.95, while the difference
is thresholded at 0.5. First row: Rhyme judgment dataset (subject
4, pseudo words contrast), Second row: Mixed gambles dataset (subject
3, parametric loss contrast). For subjects with one or several motion
spikes, the heteroscedastic and the homoscedastic PPMs differ for
a number of voxels. The reason for this is that the homoscedastic
model overestimates the constant variance term, due to time points
corresponding to motion spikes. The heteroscedastic model instead
incorporates the head motion parameters, or the temporal derivative
of them, to model these variance increases, and can thereby detect
more brain activity. \label{fig:firstlevelresults}}
\end{figure*}

{\color{black} 

\subsubsection{Sensitivity analysis}

To investigate the importance of the prior settings, the analysis of the rhyme judgment dataset was repeated for the following prior settings. 

Default:  $\tau_{\beta}=\tau_{\gamma}=10$, $\tau_{\rho}=1$, $r=0.5$, $\zeta=1$,

Analysis 2:  $\tau_{\beta}=\tau_{\gamma}=10$, $\tau_{\rho}=0.5$, $r=0.5$, $\zeta=1$,

Analysis 3:  $\tau_{\beta}=\tau_{\gamma}=10$, $\tau_{\rho}=1$, $r=0.5$, $\zeta=0.5$,

Analysis 4:  $\tau_{\beta}=\tau_{\gamma}=10$, $\tau_{\rho}=0.5$, $r=0.5$, $\zeta=0.5$,

Analysis 5: $\tau_{\beta}=\tau_{\gamma}=5$, $\tau_{\rho}=1$, $r=0.5$, $\zeta=1$,

Figure~\ref{fig:firstlevelsensitivity} shows the resulting homoscedastic and heteroscedastic PPMs for subject 4, which had the largest number of motion spikes. Lowering the prior variances $\tau_{\beta}$ and $\tau_{\gamma}$ leads to a clear decrease in detected brain activity, while the parameters for the noise process ($\tau_{\rho}$, $r$, and $\zeta$) have a small effect on the detected brain activity. 

\begin{figure*}
\hspace{0.8cm}\begin{huge}Hetero\hspace{4.5cm}Homo\hspace{3.2cm}Hetero
- Homo\end{huge}\vspace{0.5cm}

\begin{minipage}[t]{0.3\textwidth}%
\includegraphics[scale=0.6]{rhymejudgment_subject4_regressor2_heteroppm}%
\end{minipage}\hfill{}%
\begin{minipage}[t]{0.3\textwidth}%
\includegraphics[scale=0.6]{rhymejudgment_subject4_regressor2_homoppm}%
\end{minipage}\hfill{}%
\begin{minipage}[t]{0.3\textwidth}%
\includegraphics[scale=0.6]{rhymejudgment_subject4_regressor2_heteroppm_minus_homoppm}%
\end{minipage}\medskip{}
\hfill{}%

\begin{minipage}[t]{0.3\textwidth}%
\includegraphics[scale=0.6]{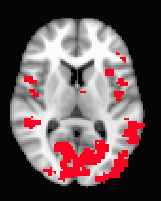}%
\end{minipage}\hfill{}%
\begin{minipage}[t]{0.3\textwidth}%
\includegraphics[scale=0.6]{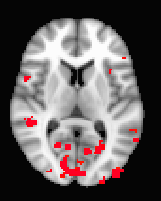}%
\end{minipage}\hfill{}%
\begin{minipage}[t]{0.3\textwidth}%
\includegraphics[scale=0.6]{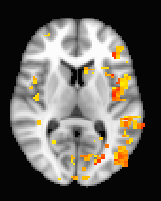}%
\end{minipage}\medskip{}
\hfill{}%

\begin{minipage}[t]{0.3\textwidth}%
\includegraphics[scale=0.6]{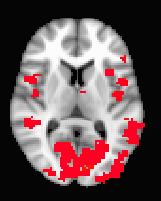}%
\end{minipage}\hfill{}%
\begin{minipage}[t]{0.3\textwidth}%
\includegraphics[scale=0.6]{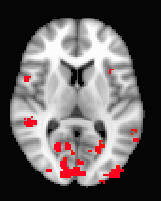}%
\end{minipage}\hfill{}%
\begin{minipage}[t]{0.3\textwidth}%
\includegraphics[scale=0.6]{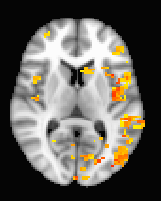}%
\end{minipage}\medskip{}
\hfill{}%

\begin{minipage}[t]{0.3\textwidth}%
\includegraphics[scale=0.6]{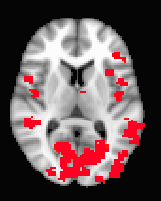}%
\end{minipage}\hfill{}%
\begin{minipage}[t]{0.3\textwidth}%
\includegraphics[scale=0.6]{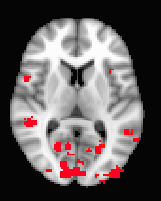}%
\end{minipage}\hfill{}%
\begin{minipage}[t]{0.3\textwidth}%
\includegraphics[scale=0.6]{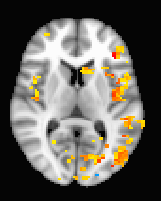}%
\end{minipage}\medskip{}
\hfill{}%

\begin{minipage}[t]{0.3\textwidth}%
\includegraphics[scale=0.6]{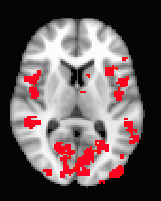}%
\end{minipage}\hfill{}%
\begin{minipage}[t]{0.3\textwidth}%
\includegraphics[scale=0.6]{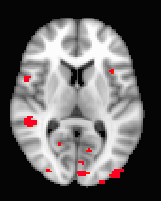}%
\end{minipage}\hfill{}%
\begin{minipage}[t]{0.3\textwidth}%
\includegraphics[scale=0.6]{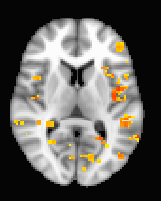}%
\end{minipage}\caption{{\color{black}Single subject posterior probability maps (PPMs) for the rhyme judgment dataset (subject
4, pseudo words contrast). From left to right: PPM for the heteroscedastic model, PPM for the homoscedastic model, PPM hetero - PPM homo. The
hetero and the homo PPMs are thresholded at Pr = 0.95, while the difference is thresholded at 0.5. First row: default prior parameters, $\tau_{\beta}=\tau_{\gamma}=10$, $\tau_{\rho}=1$, $r=0.5$, $\zeta=1$, Second row: $\tau_{\beta}=\tau_{\gamma}=10$, $\tau_{\rho}=0.5$, $r=0.5$, $\zeta=1$, Third row: $\tau_{\beta}=\tau_{\gamma}=10$, $\tau_{\rho}=1$, $r=0.5$, $\zeta=0.5$, Fourth row: $\tau_{\beta}=\tau_{\gamma}=10$, $\tau_{\rho}=0.5$, $r=0.5$, $\zeta=0.5$, Fifth row: $\tau_{\beta}=\tau_{\gamma}=5$, $\tau_{\rho}=1$, $r=0.5$, $\zeta=1$. }  \label{fig:firstlevelsensitivity}}
\end{figure*}

\subsubsection{Effect of updating $\pi_{\beta}$ and  $\pi_{\gamma}$}

To investigate the effect of updating $\pi_{\beta}$ and  $\pi_{\gamma}$ in every MCMC iteration, compared to using fix values, the analysis of the rhyme judgment dataset was repeated with and without updating the inclusion parameters. Figure~\ref{fig:firstlevelupdateinclusion} shows the resulting homoscedastic and heteroscedastic PPMs for subject 4. Updating the inclusion parameters leads to lower posterior probabilities for the activity covariates, but the difference between the heteroscedastic and homoscedastic models is still rather large. }

\begin{figure*}
\hspace{0.8cm}\begin{huge}Hetero\hspace{4.25cm}Homo\hspace{3.2cm}Hetero
- Homo\end{huge}\vspace{0.5cm}

\begin{minipage}[t]{0.3\textwidth}%
\includegraphics[scale=0.55]{rhymejudgment_subject4_regressor2_heteroppm}%
\end{minipage}\hfill{}%
\begin{minipage}[t]{0.3\textwidth}%
\includegraphics[scale=0.55]{rhymejudgment_subject4_regressor2_homoppm}%
\end{minipage}\hfill{}%
\begin{minipage}[t]{0.3\textwidth}%
\includegraphics[scale=0.55]{rhymejudgment_subject4_regressor2_heteroppm_minus_homoppm}%
\end{minipage}\medskip{}
\hfill{}%

\begin{minipage}[t]{0.3\textwidth}%
\includegraphics[scale=0.55]{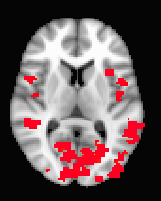}%
\end{minipage}\hfill{}%
\begin{minipage}[t]{0.3\textwidth}%
\includegraphics[scale=0.55]{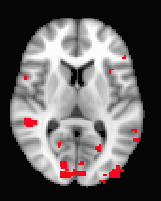}%
\end{minipage}\hfill{}%
\begin{minipage}[t]{0.3\textwidth}%
\includegraphics[scale=0.55]{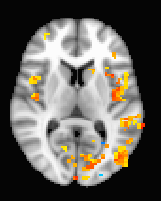}%
\end{minipage}\caption{{\color{black}Single subject posterior probability maps (PPMs) for the rhyme judgment dataset (subject
4, pseudo words contrast). From left to right: PPM for the heteroscedastic model, PPM for the homoscedastic model, PPM hetero - PPM homo. The
hetero and the homo PPMs are thresholded at Pr = 0.95, while the difference is thresholded at 0.5. First row: the inclusion parameters $\pi_{\beta}$ and  $\pi_{\gamma}$ are fixated at 0.5, Second row: the inclusion parameters $\pi_{\beta}$ and  $\pi_{\gamma}$ are updated in every MCMC iteration.}  \label{fig:firstlevelupdateinclusion}}
\end{figure*}

\subsubsection{Convergence \& efficiency of MCMC}

The MCMC convergence is in general excellent; the acceptance probabilities
for the variance covariates are 85.4\% $\pm$ 5.1\% for the rhyme
judgment dataset, 89\% $\pm$ 1.9\% for the living nonliving dataset
and 87\% $\pm$ 7.1\% for the mixed gambles dataset (standard deviation
calculated over subjects). {\color{black} Trace plots are normally used to demonstrate convergence of MCMC chains, but the large number of voxels and covariates make such visual investigations difficult. For a single subject with 10,000 voxels in gray matter, the total number of trace plots would be 440,000 (representing 20 covariates for mean and variance and four AR parameters).} The efficiency of the MCMC chain in each
voxel was {\color{black} therefore instead} investigated by calculating the inefficiency factor {\color{black} (also known as the integrated autocorrelation time~\citep{liu2008monte} defined as $1 + 2\sum_{i = 1}^{\infty} r_i$, where $r_i$ is the ith autocorrelation of the MCMC draws for a given parameter)} for each covariate for the mean and the variance, as well as for the four AR parameters. Since it is hard to estimate the inefficiency factor for variables with a low posterior inclusion probability (IPr), the
inefficiency factor was only estimated if the IPr was larger than
0.3. To carefully investigate the MCMC efficiency in every voxel is
difficult, due to the large number of voxels and covariates. An inefficiency
factor of 1 is ideal, but very seldom achieved in practice. Inefficiency
factors less than 10 - 20 are normally considered as acceptable. Tables~\ref{table:ineffvoxelsbeta},~\ref{table:ineffvoxelsgamma}
and~\ref{table:ineffvoxelsrho} therefore state the proportion of
included voxels (IPr \textgreater{} 0.3) where the inefficiency factor
is larger than 10, for the mean covariates $\left(\boldsymbol{\beta}\right)$,
the variance covariates $\left(\boldsymbol{\gamma}\right)$, and the auto regressive
parameters $\left(\boldsymbol{\rho}\right)$, respectively. The efficiency is in
general high for both the mean and the variance covariates; only a
few voxels have inefficiency factors larger than 10. The efficiency
is in general lower for the auto regressive parameters, which has
two explanations. First, the stationarity restriction enforces the
parameters to a certain region, and if a parameter is repeatedly close
to the boundary the sampling efficiency will be low. Second, in some
voxels the algorithm finds a new mode after a subset of all the draws,
which indicates that the chain has not converged. Considering the
fact that 1,000 draws are already used for burnin, and that the processing
time is 10 - 40 hours per subject, increasing the number of burnin
draws even further is not a realistic option.

\subsubsection{Group analysis}

Group analyses were performed using the full posterior of the task-related
covariates from each subject (1,000 draws). To keep things simple,
we perform each group analysis by computing the posterior for the
sample mean: $\beta_{group}=N^{-1}\sum_{r=1}^{N}\beta^{(r)}$, where
$\beta^{(r)}$ is the scalar activity coefficient for the $r$th subject
in the sample, and $N$ is the number of subjects. For each draw,
the mean brain activity over subjects was calculated, to form the
posterior of the mean group activity, $\beta_{group}$. In a second
group analysis, each subject was weighted with the inverse posterior
standard deviation, i.e. $\beta_{group}=N^{-1}\sum_{r=1}^{N}\beta^{(r)}/std(\beta^{(r)})$.
Figure~\ref{fig:groupresults} shows hetero and homo group mean PPMs
(unweighted and weighted) for the rhyme judgment dataset, minimal
differences were found for the other two datasets. The difference
between the two models is slightly larger for the weighted group analysis,
which is natural as the GLMH approach mainly affects the variance
of the posterior. The effect of using a heteroscedastic model would
clearly be stronger at the group level if many subjects (e.g. children)
in the group exhibit motion spikes.

\clearpage{}\newpage{}

\begin{figure}
\centering\includegraphics[scale=0.45]{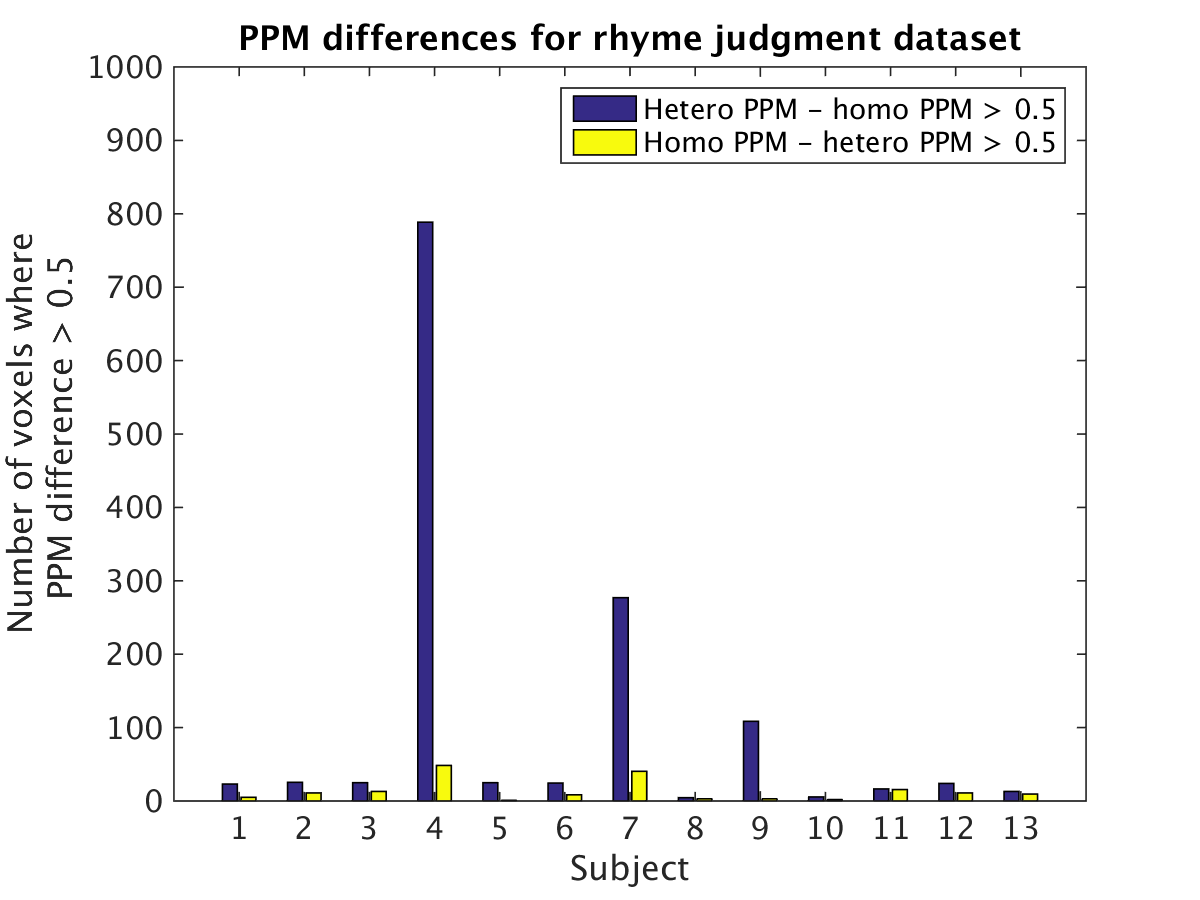}

\caption{Number of gray matter voxels where the difference between the heteroscedastic
PPM and homoscedastic PPM is larger than 0.5, for the rhyme judgment
dataset.The bars represent the average over all activity covariates.
\label{fig:ppmdiff1}}
\end{figure}

\begin{figure}
\centering\includegraphics[scale=0.45]{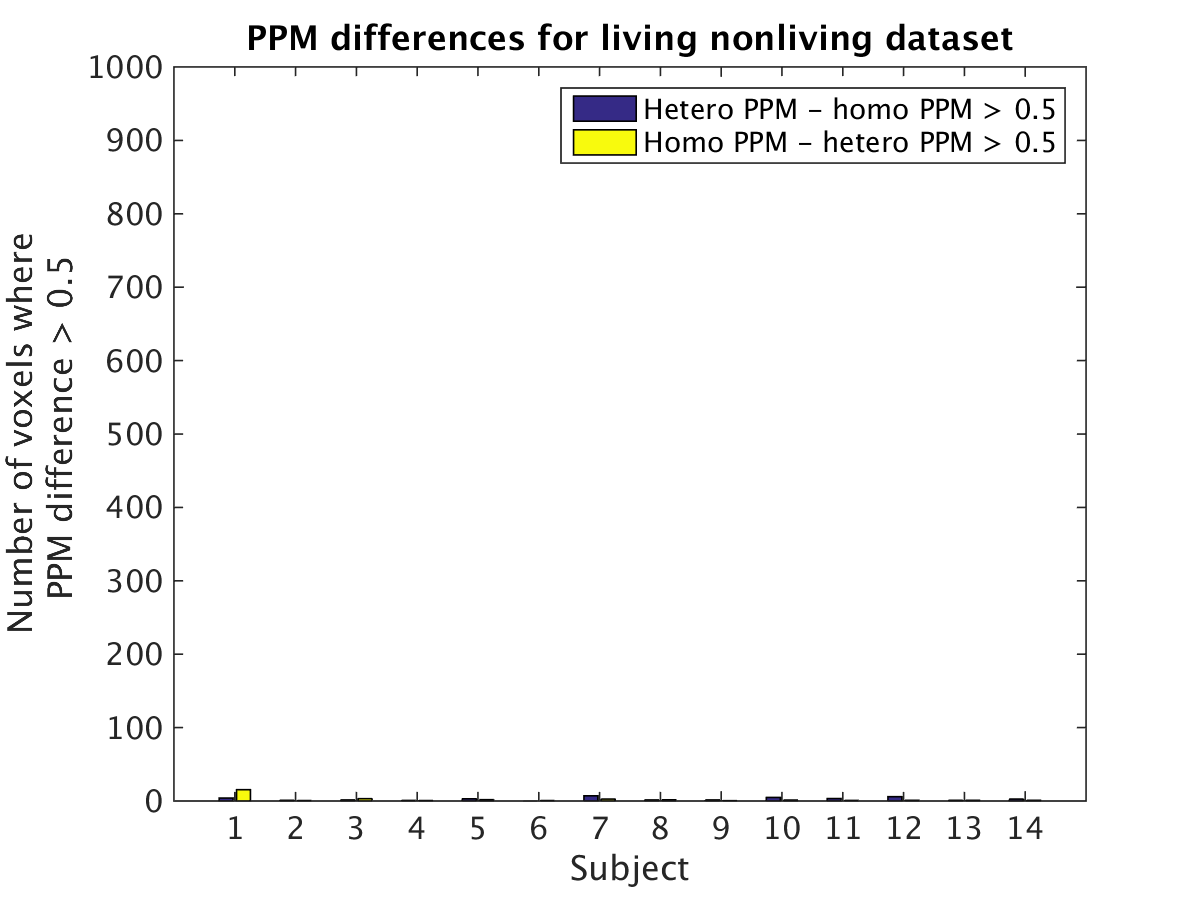}

\caption{Number of gray matter voxels where the difference between the heteroscedastic
PPM and homoscedastic PPM is larger than 0.5, for the living nonliving
dataset. The bars represent the average over all activity covariates.\label{fig:ppmdiff2}}
\end{figure}

\begin{figure}
\centering\includegraphics[scale=0.45]{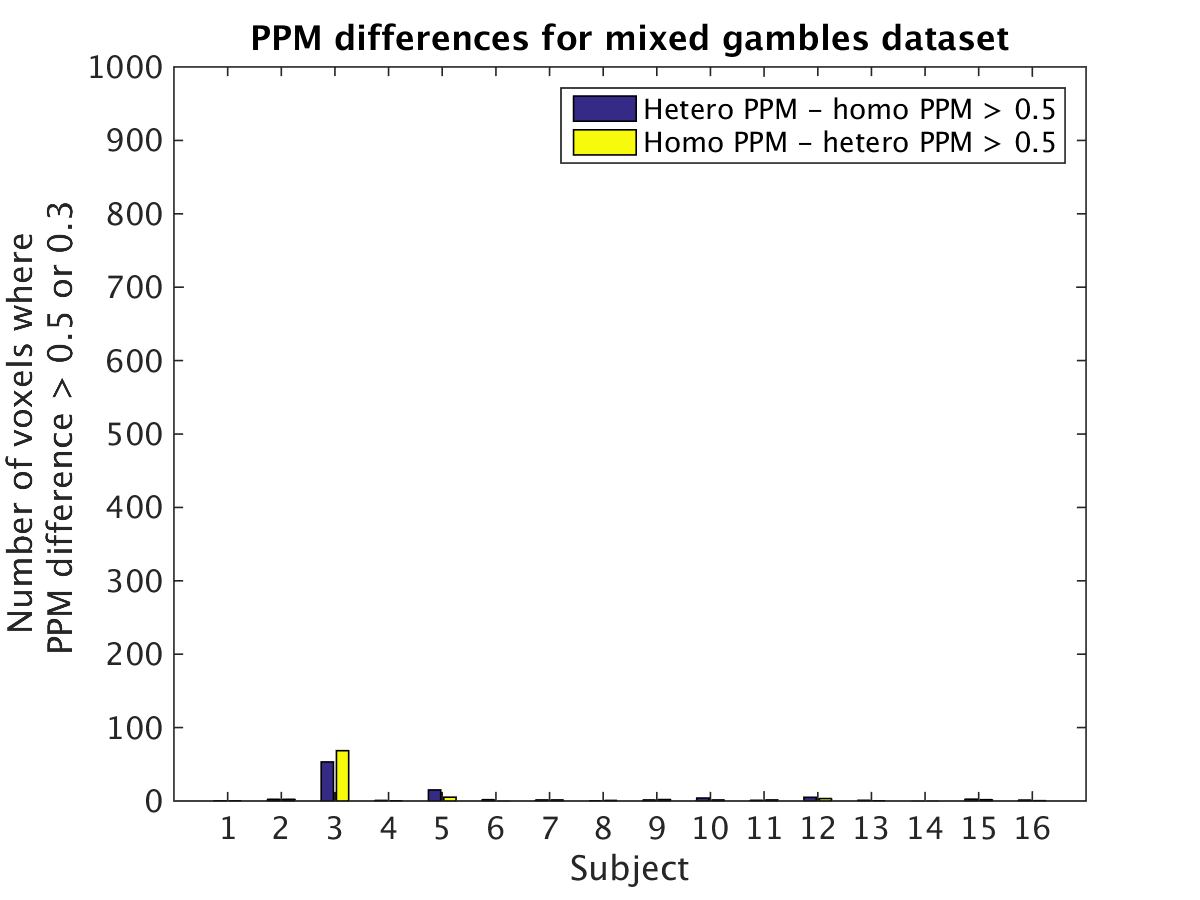}

\caption{Number of gray matter voxels where the difference between the heteroscedastic
PPM and homoscedastic PPM is larger than 0.5, for the mixed gambles
dataset. The bars represent the average over all activity covariates.
\label{fig:ppmdiff3} }
\end{figure}

\clearpage

\begin{figure}
\centering\includegraphics[scale=0.45]{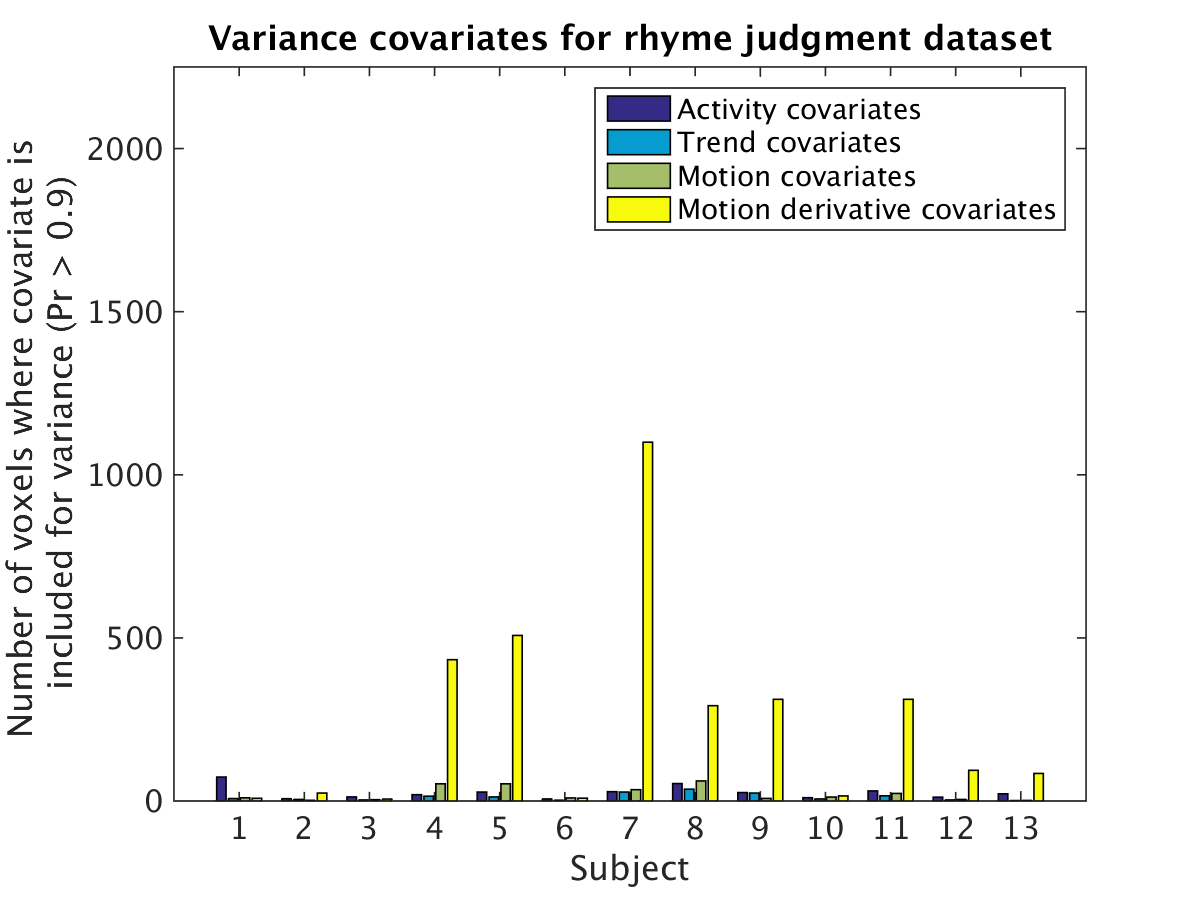}

\caption{The use of variance covariates for the rhyme judgment dataset. Each
bar represents the mean number of gray matter voxels, for each type
of covariate (activity, trends, motion, motion derivative), for which
the covariate is included to model the variance (posterior inclusion
probability larger than 0.9). For subjects with motion spikes, one
or several motion derivative covariates are used to model the heteroscedastic
variance for a large number of voxels. The mean number of gray matter
voxels is 15,600. \label{fig:varcovariates1}}
\end{figure}

\begin{figure}
\centering\includegraphics[scale=0.45]{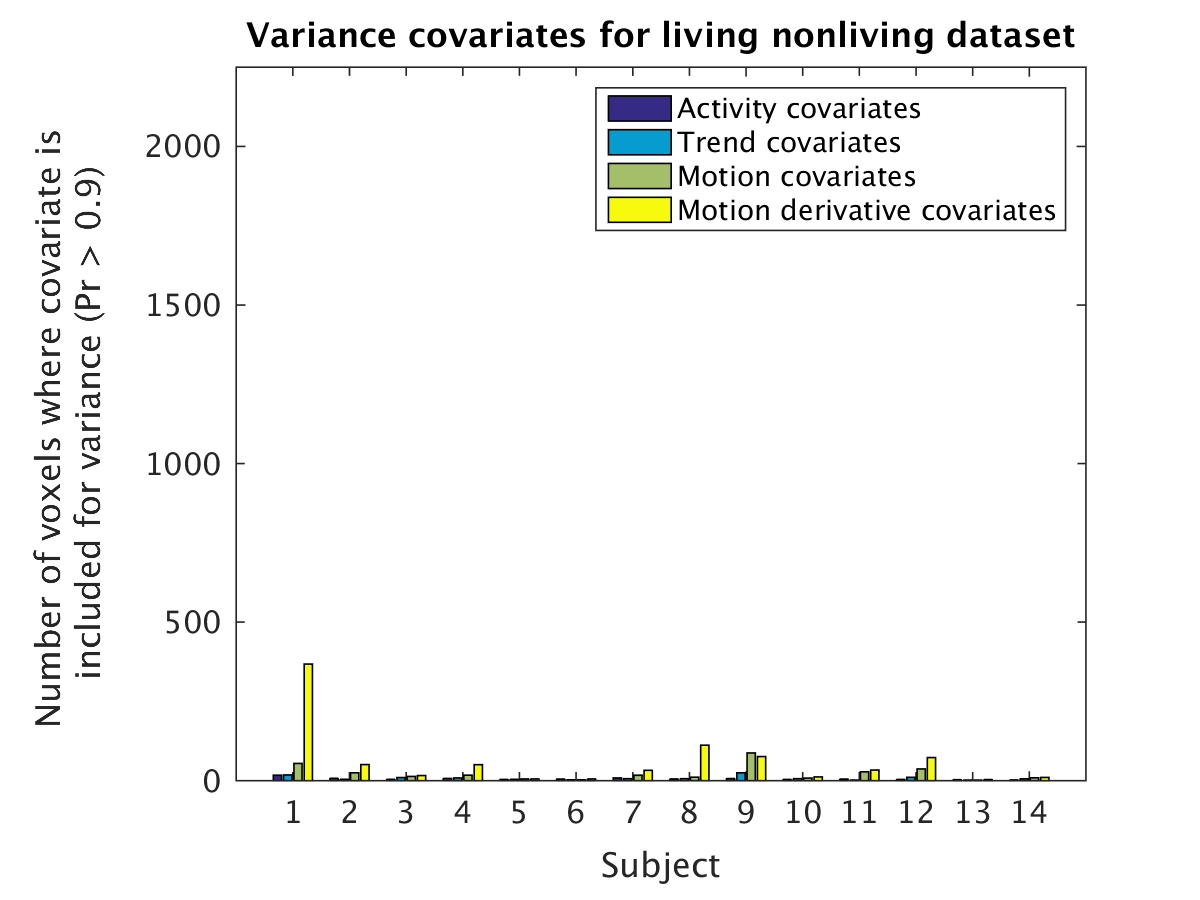}

\caption{The use of variance covariates for the living nonliving dataset. Each
bar represents the mean number of gray matter voxels, for each type
of covariate (activity, trends, motion, motion derivative), for which
the covariate is included to model the variance (posterior inclusion
probability larger than 0.9). This dataset contains very few motion
spikes, which explains why so few covariates are included in the variance.
The mean number of gray matter voxels is 13,000. \label{fig:varcovariates2}}
\end{figure}

\begin{figure}
\centering\includegraphics[scale=0.45]{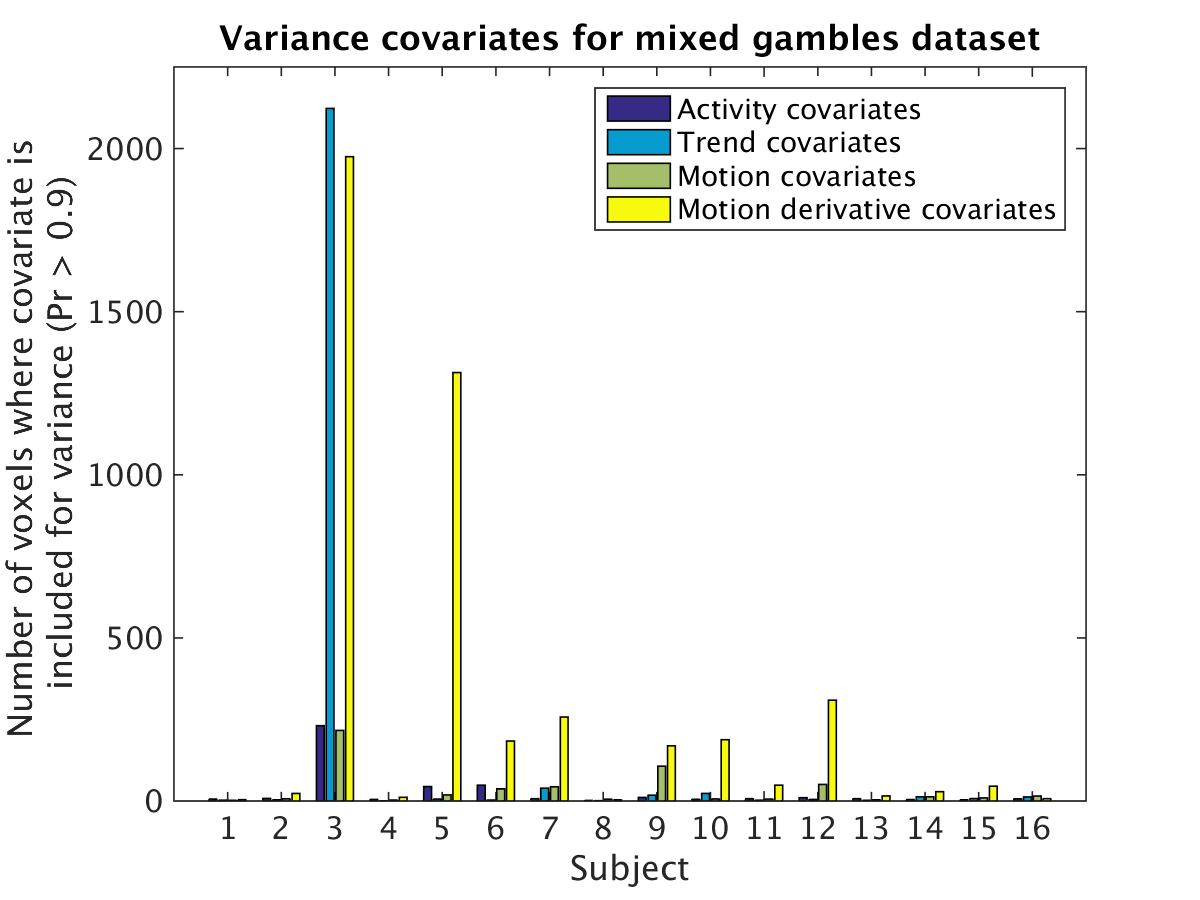}

\caption{The use of variance covariates for the mixed gambles dataset. Each
bar represents the mean number of gray matter voxels, for each type
of covariate (activity, trends, motion, motion derivative), for which
the covariate is included to model the variance (posterior inclusion
probability larger than 0.9). For subjects with motion spikes, one
or several motion derivative covariates are used to model the heteroscedastic
variance for a large number of voxels. The mean number of gray matter
voxels is 15,500. \label{fig:varcovariates3}}
\end{figure}

\clearpage

\begin{figure*}
\centering\includegraphics[scale=0.7]{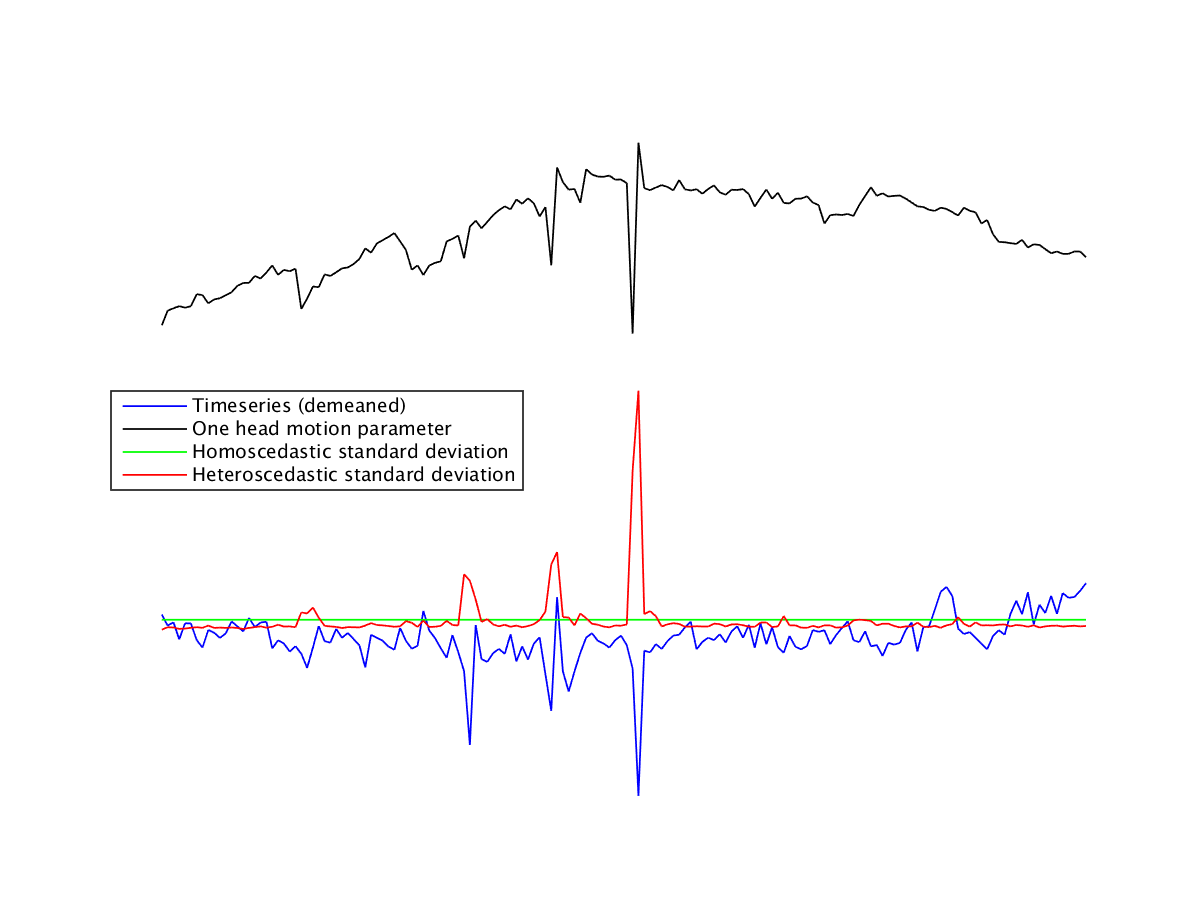}

\caption{A comparison between the estimated homoscedastic and heteroscedastic
standard deviation for one time series. The heteroscedastic standard
deviation is much higher for the motion spikes, while it is lower
for time points with little head motion. For this reason, the heteroscedastic
model can automatically downweight time points close to motion spikes,
and detect more brain activity by not over estimating the standard
deviation for time points with little head motion.. \label{fig:stdcomparison} }
\end{figure*}

\begin{table*}[htb] 
\scriptsize 
\caption{Proportion of voxels with an inefficiency factor larger than 10 for the mean covariates ($\boldsymbol{\beta}$), for the different datasets. The covariates in the design matrix have been grouped together to different types, and the numbers in the table represent the average over covariates (of each type) and subjects. The standard deviation was calculated over subjects.} 
\begin{center} 
\begin{tabular}{|c|c|c|c|c|} 
\hline  
\textbf{\normalsize Dataset / Covariate type}  & \textbf{\normalsize Activity}  & \textbf{\normalsize Time trends} & \textbf{\normalsize Motion parameters (MP)} & \textbf{\normalsize Derivative of MP} \\[0.2ex] 
\hline 
\normalsize Rhyme judgment               &  3.7\%  $\pm$ 1.9\%      & 1.8\%  $\pm$ 1.0\%   &  1.9\%  $\pm$ 0.9\%   & 2.0\%  $\pm$ 1.1\%       \\ 
\normalsize Living nonliving decision    &  2.0\%  $\pm$ 1.1\%      & 1.7\%  $\pm$ 1.2\%   &  1.5\%  $\pm$ 0.9\%   & 2.2\%  $\pm$ 1.3\%     \\ 
\normalsize Mixed gambles task           &  3.0\%  $\pm$ 1.8\%      & 3.0\%  $\pm$ 1.9\%   &  2.9\%  $\pm$ 1.7\%   & 3.4\%  $\pm$ 2.1\%   \\ 
\hline 
\end{tabular} 
\end{center} 
\label{table:ineffvoxelsbeta} 
\end{table*}

\begin{table*}[htb] 
\scriptsize 
\caption{Proportion of voxels with an inefficiency factor larger than 10 for the variance covariates ($\gamma$), for the different datasets. The covariates in the design matrix have been grouped together to different types, and the numbers in the table represent the average over covariates (of each type) and subjects. The standard deviation was calculated over subjects.} 
\begin{center} 
\begin{tabular}{|c|c|c|c|c|} 
\hline  
\textbf{\normalsize Dataset / Covariate type}  & \textbf{\normalsize Activity}  & \textbf{\normalsize Time trends} & \textbf{\normalsize Motion parameters (MP)} & \textbf{\normalsize Derivative of MP} \\[0.2ex] 
\hline 
\normalsize Rhyme judgment               &  1.3\% $\pm$ 0.8\%     &   1.0\%  $\pm$ 0.4\%   & 1.0\%  $\pm$ 0.3\%     & 0.8\%  $\pm$ 0.2\%      \\ 
\normalsize Living nonliving decision    &  1.2\% $\pm$ 0.5\%     &   1.3\%  $\pm$ 0.5\%   & 1.9\%  $\pm$ 0.5\%     & 1.4\%  $\pm$ 0.3\%      \\ 
\normalsize Mixed gambles task           &  1.8\% $\pm$ 0.6\%     &   1.8\%  $\pm$ 0.7\%   & 2.1\%  $\pm$ 0.7\%     & 1.7\%  $\pm$ 0.5\%       \\ 
\hline 
\end{tabular} 
\end{center} 
\label{table:ineffvoxelsgamma} 
\end{table*}

\begin{table*}[htb] 
\scriptsize 
\caption{Proportion of voxels with an inefficiency factor larger than 10 for the auto correlation parameters ($\rho$), for the different datasets. The standard deviation was calculated over subjects.} 
\begin{center} 
\begin{tabular}{|c|c|c|c|c|} 
\hline  
\textbf{\normalsize Dataset / AR parameter}  & \textbf{\normalsize AR 1}  & \textbf{\normalsize AR 2} & \textbf{\normalsize AR 3} & \textbf{\normalsize AR 4} \\[0.2ex] 
\hline 
\normalsize Rhyme judgment               & 20.7\%  $\pm$ 3.0\%     &  10.7\%   $\pm$ 2.4\%   & 0.8\%  $\pm$ 0.7\%    & 0.2\%  $\pm$ 0.3\%      \\ 
\normalsize Living nonliving decision    & 12.2\%  $\pm$ 3.5\%     &  10.3\%   $\pm$ 2.9\%   & 1.2\%  $\pm$ 0.9\%    & 0.1\% $\pm$ 0.2\%      \\ 
\normalsize Mixed gambles task           & 11.1\%  $\pm$ 3.5\%     &  10.6\%   $\pm$ 2.7\%   & 2.3\%  $\pm$ 1.8\%    & 0.4\%  $\pm$ 0.6\%     \\
\hline 
\end{tabular} 
\end{center} 
\label{table:ineffvoxelsrho} 
\end{table*}

\begin{figure*}
\hspace{1.8cm}\begin{huge}Hetero\hspace{4.5cm}Homo\hspace{3.5cm}Hetero
- Homo\end{huge}\vspace{0.5cm}

\begin{minipage}[t]{0.3\textwidth}%
\includegraphics{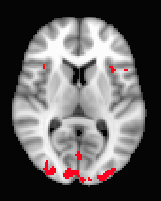}%
\end{minipage}\hfill{}%
\begin{minipage}[t]{0.3\textwidth}%
\includegraphics{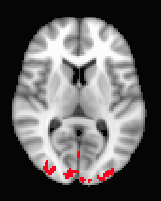}%
\end{minipage}\hfill{}%
\begin{minipage}[t]{0.3\textwidth}%
\includegraphics{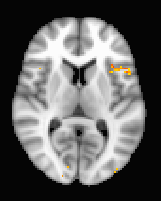}%
\end{minipage}

\begin{minipage}[t]{0.3\textwidth}%
\includegraphics{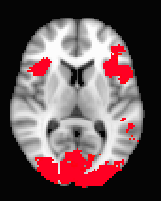}%
\end{minipage}\hfill{}%
\begin{minipage}[t]{0.3\textwidth}%
\includegraphics{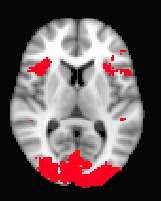}%
\end{minipage}\hfill{}%
\begin{minipage}[t]{0.3\textwidth}%
\includegraphics{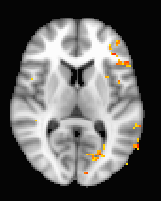}%
\end{minipage}

\caption{Group level posterior probability maps (PPMs) for the rhyme judgment
dataset (contrast pseudo words). From left to right: PPM for the heteroscedastic
model, PPM for the homoscedastic model, PPM hetero - PPM homo. The
hetero and the homo PPMs are thresholded at Pr = 0.95, while the difference
is thresholded at 0.5. Top row: group activity calculated without
any subject specific weights. Bottom row: group activity calculated
by weighting each subject with the inverse standard deviation. \label{fig:groupresults}}
\end{figure*}

\clearpage{}

\section{Discussion}

We have presented a Bayesian heteroscedastic GLM for single subject
fMRI analysis. The heteroscedastic GLM takes into consideration the
fact that the variance is inflated for time points with a high degree
of head motion, and thus provides more sensitive results, compared
to its homoscedastic counterpart. Instead of discarding data with
too much head motion, or applying different scrubbing or censoring
techniques \citep{Satterthwaite,Power2014,siegel}, our heteroscedastic
GLM automatically downweights the affected time points, and propagates
the uncertainty to the group analysis by saving the full posterior.
For the rhyme judgment dataset and the mixed gambles dataset, the
temporal derivative of the head motion parameters are included as
variance covariates for a large number of voxels. For heteroscedastic
voxels in active brain areas, the difference between the homoscedastic
PPM and the heteroscedastic PPM can be substantial. There will only
be a sizeable PPM difference if the voxel belongs to an active brain
area, and contains noise where the degree of heteroscedasticity is
sufficiently high (see Figures~\ref{fig:simulationresults1} -~\ref{fig:simulationresults4}).
The difference between the two models is small for the living/nonliving
dataset, mainly because that dataset contains very few motion spikes.
This illustrates that our algorithm can be applied to any dataset,
as using the heteroscedastic approach does not lead to a lower sensitivity
when there are no motion spikes present. 

\subsection{MCMC vs Variational Bayes}

A drawback of using MCMC is the computational complexity; it takes
10 - 40 hours (depending on the number of covariates) to analyze a
single subject using the heteroscedastic model, with a {\color{black} single} Intel Core
i7 4790K CPU {\color{black} with 4 physical cores (8 cores due to hyper threading)  and 32 GB of RAM}. One alternative is to use variational Bayes
(VB), where a few iterations is normally sufficient to obtain a point
estimate of the posterior \citep{Penny2003}. It is, however, much
harder to perform variable selection within VB, and variable selection
is necessary in our case since 18 - 21 covariates are used for the
mean as well as for the variance. Without variable selection the model
would contain too many parameters, compared to the number of time
points in a typical fMRI dataset, which would result in poor estimates.
Another problem with VB is that the posterior standard deviation is
often underestimated.

In theory, the proposed algorithm can run on a graphics processing
unit (GPU), which can analyze some 30,000 voxels in parallel \citep{Eklund2013,Eklund2014}.
The pre-whitening step in each MCMC iteration is problematic from
a GPU perspective, as a pre-whitened design matrix needs to be stored
in each voxel / GPU thread. For 20 covariates and 200 time points,
the design matrix requires 4,000 floats for storage. Modern Nvidia
GPUs can, however, only store 255 floats per thread. 

\subsection{GLMH vs weighted least squares}

To make a fair comparison between our heteroscedastic model and the
WLS approach proposed by \citet{Diedrichsen2005} is difficult, as
we use Bayesian inference. Nevertheless, the WLS approach seems to
work well as long as the same heteroscedastic noise is present in
all voxels, but fails to detect activity when the heteroscedastic
noise is only present in 30\% of the voxels. \citet{Diedrichsen2005}
argue that the same weight should be used for all voxels, our results
for real fMRI data (Figures~\ref{fig:varcovariates1} -~\ref{fig:varcovariates3})
instead suggest that only a fraction of voxels have heteroscedastic
noise. For some 13,000 - 15,600 voxels in gray matter, the derivative
of the head motion parameters are included as covariates for the variance
for 300 - 2,000 voxels (for subjects with motion spikes). Note that
these numbers represent the average over each covariate type, meaning
that if one of the six motion covariates is included for 12,000 voxels,
the average over all six covariates will be 2,000 voxels. 

The main drawback of the WLS approach is that it requires estimation
of $T$ weights from $T$ time points, which results in extremely
variable estimates unless the weights are averaged over many voxels.
Our heteroscedastic GLM instead models the variance using a regression
approach. Through the use of variable selection, a heteroscedastic
model can be estimated independently in each voxel, even if the number
of covariates is large.

\subsection{Multiple comparisons}

In contrast to frequentistic statistics, there is no consensus in
the fMRI field regarding if and how to correct for multiple comparisons
for PPMs. In this paper we have mainly focused on looking at differences
between the heteroscedastic and the homoscedastic models, for voxel
inference. It is not obvious how to use Bayesian techniques for cluster
inference \citep{eklundpnas}, which for frequentistic statistics
has a higher statistical power. One possible approach is to use theory
on excursion sets \citep{bolin}, to work with the joint PPM instead
of marginal PPMs. Such an approach, however, requires a spatially
dependent posterior, while we independently estimate one posterior
for each voxel. {\color{black} One ad-hoc approach is to calculate a Bayesian t- or z-score for
each voxel, and then apply existing frequentistic approaches for multiple comparison correction (e.g. Gaussian random field theory). This approach
is for example used in the FSL software~\citep{flame}.} 

\subsection{Future work}

We have here only demonstrated the use of the heteroscedastic GLM
for brain activity estimation, but it can also be used for estimating
functional connectivity; for example by using a seed time series as
a covariate in the design matrix. Although not investigated in this
work, it is also possible to include additional covariates that may
affect the variance, such as the global mean \citep{Power2016} or
recordings of breathing and pulse \citep{glover}. Future work will
also focus on adding a spatial model \citep{penny2005,siden}, instead
of analyzing each voxel independently.

\section*{Acknowledgement}

\vspace{-0.3cm}

This work was financed by the Swedish Research council, grant 2013-5229
(``Statistical analysis of fMRI data''), and by the Information
Technology for European Advancement (ITEA) 3 Project BENEFIT (better
effectiveness and efficiency by measuring and modelling of interventional
therapy). This research was also supported in part by NIH grants R01
EB016061 and P41 EB015909 from the National Institute of Biomedical
Imaging. We thank Russ Poldrack and his colleagues for starting the
OpenfMRI Project (supported by National Science Foundation Grant OCI-1131441)
and all of the researchers who have shared their task-based data.

\vspace{-0.3cm}

\bibliographystyle{apalike}
\bibliography{fMRI}

%\clearpage{}

\section{Appendix A - Implementation \label{Appendix A}}

Our heteroscedastic GLM can be launched from a Linux terminal as\medskip{}

HeteroGLM fmri.nii.gz -designfiles activitycovariates.txt 

-gammacovariates gammacovariates.txt 

-ontrialbeta trialbeta.txt -ontrialgamma trialgamma.txt

-ontrialrho trialrho.txt -mask mask.nii.gz 

-regressmotion motion.txt 

-regressmotionderiv motionderiv.txt 

-draws 1000 -burnin 1000 -savefullposterior

{\color{black}-updateinclusionprob}

\medskip{}
where ``activitycovariates.txt'' states the activity covariates for
the design matrix (normally only used for the mean), ``gammacovariates.txt''
states the covariates being used to model the variance, ``ontrialbeta.txt''
states covariates for which variable selection is performed for the
mean, ``ontrialgamma.txt'' states covariates for which variable selection
is performed for the variance and ``ontrialrho.txt'' states variable
selection parameters for the autocorrelation. {\color{black} The option ``updateinclusionprob'' turns on updating the inclusion probabilities $\pi_{\beta}$ and $\pi_{\gamma}$ in every MCMC iteration.} Covariates for intercept
and time trends are automatically added internally. A homoscedastic
GLM can easily be obtained as a special case, using only a single
covariate (the intercept) for the variance. The following nifti files
are created; posterior mean of beta and Ibeta (for each covariate),
posterior mean of gamma and Igamma (for each covariate), posterior
mean of rho and Irho (for each AR parameter), and PPMs for each activity
covariate. The full posterior of all beta, gamma and rho parameters
can also be saved as nifti files.

\section{Appendix B - MCMC Details \label{Appendix B}}

\subsection*{Variable selection by MCMC in the linear regression model}

Let us assume a general multivariate prior $\boldsymbol{\beta}_{\mathcal{I}}\vert\mathcal{I}\sim N\left(\boldsymbol{\mu},\Omega_{I}\right)$.
Now,
\begin{multline*}
p(\boldsymbol{\beta},\mathcal{I}\vert\mathbf{y},\mathbf{X},\mathbf{Z},\cdot)\propto p(\mathbf{y}\vert\boldsymbol{\beta},\mathcal{I},\mathbf{X},\mathbf{Z})p(\boldsymbol{\beta}\vert\mathcal{I})p(\mathcal{I})\\
\propto\exp\left(-\frac{1}{2}\left(\mathbf{y}-\mathbf{X}_{\mathcal{I}}^{T}\boldsymbol{\beta}_{\mathcal{I}}\right)^{T}\left(\mathbf{y}-\mathbf{X}_{\mathcal{I}}^{T}\boldsymbol{\beta}_{\mathcal{I}}\right)\right)\\
\times\left|2\pi\Omega_{I}\right|^{-1/2}\exp\left(-\frac{1}{2}\left(\boldsymbol{\beta}_{\mathcal{I}}-\boldsymbol{\mu}\right)^{T}\Omega_{I}^{-1}\left(\boldsymbol{\beta}_{\mathcal{I}}-\boldsymbol{\mu}\right)\right)p(\mathcal{I}),
\end{multline*}
where $\mathbf{X}_{\mathcal{I}}$ is the matrix formed by selecting
the columns of $\mathbf{X}$ corresponding to $\mathcal{I}$. The
conditional likelihood\\
 $\exp\left(-\frac{1}{2}\left(\mathbf{y}-\mathbf{X}_{\mathcal{I}}^{T}\boldsymbol{\beta}_{\mathcal{I}}\right)^{T}\left(\mathbf{y}-\mathbf{X}_{\mathcal{I}}^{T}\boldsymbol{\beta}_{\mathcal{I}}\right)\right)$
can be decomposed as
\begin{multline*}
\exp\left(-\frac{1}{2}\left[\left(\mathbf{y}-\mathbf{X}_{\mathcal{I}}\hat{\boldsymbol{\beta}}_{\mathcal{I}}\right)^{T}\left(\mathbf{y}-\mathbf{X}_{\mathcal{I}}\hat{\boldsymbol{\beta}}_{\mathcal{I}}\right)\right]\right)\\
\times\exp\left(-\frac{1}{2}\left[\left(\boldsymbol{\beta}_{\mathcal{I}}-\hat{\boldsymbol{\beta}}_{\mathcal{I}}\right)^{T}\mathbf{X}_{\mathcal{I}}^{T}\mathbf{X}_{\mathcal{I}}\left(\boldsymbol{\beta}_{\mathcal{I}}-\hat{\boldsymbol{\beta}}_{\mathcal{I}}\right)\right]\right)
\end{multline*}
Multiplying the conditional likelihood by the prior and completing
the square \foreignlanguage{swedish}{}\footnote{\selectlanguage{swedish}%
$(x-a)'A(x-a)+(x-b)'B(x-b)=(x-d)'D(x-d)+(d-a)'A(d-a)+(d-b)'B(d-b)$,
where $D=A+B$ and $d=D^{-1}(Aa+Bb).$\selectlanguage{english}%
} gives
\begin{multline*}
p(\boldsymbol{\beta},\mathcal{I}\vert\mathbf{y},\mathbf{X},\mathbf{Z},\cdot)\propto\\
c\cdot\exp\left(-\frac{1}{2}\left[\left(\boldsymbol{\beta}_{\mathcal{I}}-\tilde{\boldsymbol{\beta}}_{I}\right)^{T}\left(\mathbf{X}_{\mathcal{I}}^{T}\mathbf{X}_{\mathcal{I}}+\Omega_{I}^{-1}\right)\left(\boldsymbol{\beta}_{\mathcal{I}}-\tilde{\boldsymbol{\beta}}_{I}\right)\right]\right)\\
\times\exp\left(-\frac{1}{2}\left[\left(\tilde{\boldsymbol{\beta}}_{I}-\hat{\boldsymbol{\beta}}_{I}\right)^{T}\mathbf{X}_{\mathcal{I}}^{T}\mathbf{X}_{\mathcal{I}}\left(\tilde{\boldsymbol{\beta}}_{I}-\hat{\boldsymbol{\beta}}_{I}\right)\right]\right)\\
\times\exp\left(-\frac{1}{2}\left[\left(\tilde{\boldsymbol{\beta}}_{I}-\boldsymbol{\mu}\right)^{T}\Omega_{I}^{-1}\left(\tilde{\boldsymbol{\beta}}_{I}-\boldsymbol{\mu}\right)\right]\right)p(\mathcal{I})
\end{multline*}
where $c=\left|2\pi\Omega_{I}\right|^{-1/2}\exp\left(-\frac{1}{2}\left(\mathbf{y}-\mathbf{X}_{\mathcal{I}}\hat{\boldsymbol{\beta}}_{\mathcal{I}}\right)^{T}\left(\mathbf{y}-\mathbf{X}_{\mathcal{I}}\hat{\boldsymbol{\beta}}_{\mathcal{I}}\right)\right)$
and $\tilde{\boldsymbol{\beta}}_{I}=\left(\mathbf{X}_{\mathcal{I}}^{T}\mathbf{X}_{\mathcal{I}}+\Omega_{I}^{-1}\right)^{-1}\left(\mathbf{X}_{\mathcal{I}}^{T}\mathbf{X}_{\mathcal{I}}\hat{\boldsymbol{\beta}}+\Omega_{I}^{-1}\boldsymbol{\mu}\right)$.
This shows that
\[
\boldsymbol{\beta}_{\mathcal{I}}\vert\mathcal{I}\sim N\left(\tilde{\boldsymbol{\beta}}_{I},\left(\mathbf{X}_{\mathcal{I}}^{T}\mathbf{X}_{\mathcal{I}}+\Omega_{I}^{-1}\right)^{-1}\right).
\]
Integrating with respect to $\boldsymbol{\beta}_{\mathcal{I}}$ gives
\begin{multline*}
p(\mathcal{I}\vert\mathbf{y},\mathbf{X},\mathbf{Z},\cdot)\propto\left|\Omega_{I}\mathbf{X}_{\mathcal{I}}^{T}\mathbf{X}_{\mathcal{I}}+I\right|^{-1/2}\\
\times\exp\left(-\frac{1}{2}\left(\mathbf{y}-\mathbf{X}_{\mathcal{I}}\hat{\boldsymbol{\beta}}_{\mathcal{I}}\right)^{T}\left(\mathbf{y}-\mathbf{X}_{\mathcal{I}}\hat{\boldsymbol{\beta}}_{\mathcal{I}}\right)\right)\\
\times\exp\left(-\frac{1}{2}\left[\left(\tilde{\boldsymbol{\beta}}_{I}-\hat{\boldsymbol{\beta}}_{I}\right)^{T}\mathbf{X}_{\mathcal{I}}^{T}\mathbf{X}_{\mathcal{I}}\left(\tilde{\boldsymbol{\beta}}_{I}-\hat{\boldsymbol{\beta}}_{I}\right)\right]\right)\\
\times\exp\left(-\frac{1}{2}\left[\left(\tilde{\boldsymbol{\beta}}_{I}-\boldsymbol{\mu}\right)^{T}\Omega_{I}^{-1}\left(\tilde{\boldsymbol{\beta}}_{I}-\boldsymbol{\mu}\right)\right]\right)p(\mathcal{I}).
\end{multline*}

\end{document}